\documentclass[preprint,final,aps,prb,amsfonts,letterpaper,citeautoscript]{revtex4-1}

\usepackage{CJK} 
\usepackage{graphicx, multirow,bm, amsmath}
\usepackage{hyperref}
\usepackage{bbm}
\usepackage[usenames,dvipsnames]{color}
\usepackage[version=4]{mhchem}

\begin{document}
\begin{CJK*}{UTF8}{bsmi} 
	\title{Compressive sensing lattice dynamics. I. General formalism}

	\author{Fei Zhou(周非)} \email{fzhou@llnl.gov}
	\affiliation{Physical and Life Sciences Directorate, Lawrence Livermore National Laboratory, Livermore, California 94550, USA}
	\author{Weston Nielson} 
	\affiliation{Department of Materials Science and Engineering, University of California, Los Angeles, California 90095-1595, USA} 
	\author{Yi Xia}
	\affiliation{Department of Materials Science and Engineering, Northwestern University, Evanston, IL 60208, USA}
	\author{Vidvuds Ozoli\c{n}\v{s}}		\email{vidvuds.ozolins@yale.edu}
	\affiliation{Department of Applied Physics, Yale University, New Haven, CT 06520, USA} 
	\affiliation{Yale Energy Sciences Institute, Yale University, West Haven, CT 06516, USA} 
	
	\date{\today} 

\begin{abstract}
{\it Ab initio\/} calculations have been successfully used for evaluating lattice dynamical properties of solids within the (quasi-)harmonic approximation (i.e., assuming non-interacting phonons with infinite lifetimes), but it remains difficult to treat anharmonicity in all but the simplest compounds.  We detail a systematic information theory based approach to deriving {\it ab initio\/} anharmonic force constants: compressive sensing lattice dynamics (CSLD). 
The non-negligible terms that are necessary to reproduce the first-principles calculated interatomic forces are automatically selected by minimizing the $\ell_1$ norm (sum of absolute values) of the scaled force constants. By using efficient sampling of the configuration space using a modest number of atomic configurations with quasi-random displacements, CSLD is well suited for deriving accurate anharmonic potentials for complex multicomponent compounds with large unit cells.
We demonstrate the power and generality of CSLD by calculating the phonon lifetimes and thermal transport properties of Type-I Si clathrates.
\end{abstract}
\maketitle
\end{CJK*}

\section{Introduction}
The dynamics of atomic vibrations plays a central role in the structural, thermodynamic, and transport properties of crystalline solids at finite temperature. The quantum theory of lattice dynamics (LD) can be traced back to the pioneering Einstein \cite{Einstein1907AP180} and Debye models \cite{Debye1912AP789} for the specific heat of solids at low temperature. The modern theory of lattice dynamics \cite{Born1912PZ297,*Born1912PZ15,Born-Huang1954Dynamical} forms the basis for our quantitative understanding of lattice vibrations and their relation to macroscopic properties.  With the advent of efficient density-functional theory (DFT) based methods for calculating the potential energy surfaces (PES) of solids,  LD  provides a bridge between atomistic quantum-mechanical calculations at zero temperature and macroscopic  properties at finite temperature \cite{Baroni2001RMP515, vandeWalle2002RMP11, Fultz2010PMS247, Grimvall2012RMP945}.

Phonons are quasiparticles representing collective vibrations of the crystal lattice. Non-interacting phonons arise from the second order (harmonic) Taylor expansion of the PES and can be readily calculated with first-principles methods \cite{Baroni2001RMP515, Togo2015SM1, Wang2016CM16006}. Phonon-phonon interactions beyond the harmonic approximation give rise to many important physical phenomena related to finite phonon lifetimes and phonon frequency shifts, such as phonon scattering, lattice thermal conductivity, structural phase transformations, and ferroelectricity. 
However, first principles treatment of these effects is presently less ubiquitous, since a practical and systematic DFT-based approach to anharmonicity has proven more challenging \cite{Debernardi1995PRL1819,Broido2007APL231922, Garg2011PRL045901,Esfarjani2008PRB144112,Esfarjani2011PRB085204, Hellman2013PRB104111,  Li2012PRB174307, Li2014CPC1747,Ozolins2009PRL065702,Glensk2014PRX011018, Glensk2015PRL195901}.  In principle, the ``$2n+1$'' theorem \cite{Gonze1989PRB13120} of density functional perturbation theory (DFPT) \cite{Baroni1987PRL1861, Gonze1995PRA1096} is generally applicable and gives the higher-order anharmonicity. However, such computations are cumbersome and require specialized codes that are not widely available for $n >1$ \cite{Feng2016PRB045202}.  
Alternatively, the anharmonic terms may also in principle be calculated by finite-difference, but the task is severely hindered by the combinatorial explosion in the number of parameters with increasing order and interaction distance \cite{Togo2015SM1, Tadano2014JPCM225402}, but also by the growing numeric errors and instability associated with high order polynomials. Additionally, a Taylor expansion of the nearest neighbor interaction potential is inherently inefficient and cumbersome compared to e.g.\ a force field.

Faced with the challenge of direct computation of the higher-order anharmonic terms, various alternatives to circumventing the Taylor expansion of the PES have been developed. 
The self-consistent phonon (SCPH) approximation \cite{Werthamer1970PRB572}  describes phonon spectra at finite temperature using temperature dependent effective force constants.  A thermally averaged effective harmonic Hamiltonian can be constructed by first-principles molecular dynamics (MD) simulation \cite{Hellman2011PRB180301} or by an iterative procedure in the self-consistent {\it ab initio\/} lattice dynamics (SCAILD) method \cite{Souvatzis2008PRL095901, Souvatzis2009CMS888}. These methods have been successfully applied to study thermally renormalized phonon spectra of mechanically unstable metals (bcc Li, Ti, Zr, Hf, Sc and Y). SCPH has also been extended to calculate third-order anharmonicity in Si and FeSi \cite{Hellman2013PRB144301}. 
Errea and co-workers have developed a non-perturbative stochastic approach to strong anharmonicity by renormalizing even-order potential terms into effective harmonic interactions with temperature-dependent Boltzmann weights \cite{Errea2013PRL177002,*Errea2014PRB064302}. 
In contrast to the aforementioned methods that do not attempt to explicitly treat anharmonic PES, several recent papers have exploited crystal symmetry to represent the PES in carefully designed functional forms. For instance, Wojdel {\it et al.\/}  expanded the PES in terms of displacement differences between pairs of atoms, and explicitly considered the coupling between strain and local displacements \cite{Wojdel2013JPCM305401}. 
High-temperature free energy model of \ce{ZrH2} was proposed by Thomas and Van der Ven based on symmetry-adapted cluster expansion of lattice deformation \cite{Thomas2013PRB214111}.
Very recently, Kadkhodaei and co-workers proposed the piecewise polynomial potential partitioning (P4) method to calculate  the free energy of mechanically unstable  bcc Ti \cite{Kadkhodaei2017PRB064101}.
Last but not least, methods that take advantage of advanced machine-learning techniques to construct interatomic potentials have shown promise \cite{Bartok2010PRL136403, Behler2016JCP170901}.

Nevertheless, the Taylor expansion based approach to anharmonic lattice dynamics remains very attractive. Its generality and simple mathematical form ensure broad applicability to a wide variety of crystal structures, bonding types, and physical properties. The purpose of this work is to lift the main practical obstacle to obtaining Taylor expansion of the PES beyond the quasi-harmonic level: the numerical difficulty of calculating a large number of unknown force constants. We utilize compressive sensing (CS), a technique  developed in the field of information science for recovering sparse solutions from incomplete data \cite{Candes2008IEEESPM21}, to simultaneously determine which anharmonic terms are important and to find their values\cite{Nelson2013PRB035125,Nelson2013PRB155105}. Our approach -- compressive sensing lattice dynamics (CSLD) -- can handle large, complex unit cells and strong anharmonicity, including materials with harmonically unstable phonon modes (e.g., sixth order terms were shown to be necessary for the thermoelectric compound \ce{Cu12Sb4S13}). Since the initial short publication in Ref. \onlinecite{Zhou2014PRL185501}, the CSLD method has been used in {\it ab initio} studies of $\delta$-Pu \cite{Soderlind2015SR15958}, \ce{SrI2} \cite{Zhou2016CM16022}, \ce{CsPbCl(Br)3} \cite{Guo2017ACSEL2463}, and several thermoelectric materials \cite{He2016PRL046602,Lu2016CM1781, He2016CM2529,Xia2018APL181906,pbte2018,gete2018,Hodges2018,Shen2018JMCA24877, Kuo2019JMCA2589}.  An independent implementation by Tadano and Tsuneyuki extracted anharmonic force constants and studied temperature-dependent phonon spectra in \ce{SrTiO3} \cite{Tadano2015PRB054301} and in Eu$_{8}$Ga$_{16}$Ge$_{30}$ \cite{Tadano2018PRL105901}. The CS technique has also been used to fit interatomic potentials \cite{Seko2015PRB054113}. 

Here, we lay out the general theory and computational techniques of lattice dynamics specifically adapted for a (possibly underdetermined) linear problem in Section \ref{sec:LD}. The numerical CS method for the lattice dynamics is described in Section \ref{sec:CS}, and select examples of the efficacy of CSLD are presented in Section \ref{sec:results}. 
Part II of this two-paper set applies CSLD to efficient phonon calculations.\cite{CSLD-phonon}

\section{Theory: Lattice dynamics} \label{sec:LD}
The basic theory of lattice dynamics is well known, and comprehensive treatment can be found in classical textbooks, e.g., in Ref.~\onlinecite{Horton1974dynamical}. Here we outline the theory in an abstract, cluster-based form that is completely general in terms of expansion orders and crystal lattice symmetry, easy to keep track of, systematically improvable, and well suited for a numeric approach. This helps us manage the staggering  numerical complexity that is inevitable if one desires to account for high-order anharmonicity. More importantly, the formulation makes it possible to extend the LD model to encompass additional degrees of freedom, such as alloying, magnetism, and charge disorder.

\subsection{General formalism}
We start by Taylor-expanding the Born-Oppenheimer potential energy $E$ of a crystalline solid in atomic displacements:
\begin{equation}
\label{eq:Taylor}
E = E_{0} + \Phi_{\mathbf{a}} u_{\mathbf{a}} + \frac{\Phi_{\mathbf{ab}} }{2} u_{\mathbf{a}} u_{\mathbf{b}} + \frac{ \Phi_{\mathbf{abc}}}{3!}  u_{\mathbf{a}} u_{\mathbf{b}}u_{\mathbf{c}}+ \cdots ,
\end{equation}
where $E_{0}$ is the total energy of a reference structure with zero displacements, $\mathbf{a} \equiv \{a,i\}$ is a composite index for atom (lattice site) $a$ and Cartesian direction $i$ ($=1$--3), $u_{\mathbf{a}} =r_{a,i}- r^0_{a,i}$ is the displacement of the atomic position $\mathbf{r}_{a}$ relative to the reference position $\mathbf{r}^0_a$. The second-order expansion coefficients $\Phi_{\mathbf{ab}} \equiv \Phi_{ij}(ab) = \partial^{2} E/\partial u_{\mathbf{a}} \partial u_{\mathbf{b}}$ determine the phonon dispersion in the harmonic approximation, and $\Phi_{\mathbf{abc}} \equiv \Phi_{ijk}(abc) =  \partial^{3} E/\partial u_{\mathbf{a}} \partial u_{\mathbf{b}} \partial u_{\mathbf{c}}$ is the third-order anharmonic force constant tensor (FCT).
In general, an order-$n$ FCT is defined as
\begin{eqnarray}
\label{eq:deritative}
\Phi_{i_{1} \dots i_{n}}(a_{1} \dots a_{n}) =   \partial^{n} E/ \partial u_{\mathbf{a}_{1}} \dots \partial u_{\mathbf{a}_{n}}.
\end{eqnarray}
The linear term $\Phi_{\mathbf{a}}  $ is absent when the reference structure represents mechanical equilibrium. The Einstein summation convention over repeated indices is implied. 

Systematic calculation or fitting of the higher-order anharmonic terms is challenging due to a combinatorial explosion in the number of tensors $\Phi(a_{1} \cdots a_{n})$ with increasing order $n$ and maximum distance between the sites $\{a_{1}, \ldots, a_{n}\}$, as well as the number of elements $3^n$ in an order-$n$ tensor. To reduce the complexity and truncate Eq.~(\ref{eq:Taylor}) to a manageable form, one may rely on physical intuition, e.g.\ that the largest anharmonic terms correspond to neighboring atoms with direct chemical bonds and hence are short-ranged, while long range interactions vary slowly and can be accurately described using {\it harmonic\/} FCTs. Once the long-range Coulombic force constants have been accounted for (our prescription is given in Ref.~\onlinecite{CSLD-phonon}), the remaining interactions are expected to be short-ranged, i.e., they decay faster than the 3$^\text{rd}$ power of the interatomic distance \cite{Gonze1997PRB10355}. However, such knowledge is generally not {\it a priori\/} obvious in a complex system and can only be gained on a case-by-case basis through time-consuming cycles of model construction and cross-validation. As a result,  anharmonic FCTs can be usually calculated only for relatively simple crystals and weak anharmonicity \cite{Esfarjani2008PRB144112,Esfarjani2011PRB085204,Hellman2013PRB144301}.

Equation~(\ref{eq:Taylor}) can be written in a more convenient {\bf multi-index} notation (details in Appendix \ref{sec:multi-index}):
\begin{eqnarray}
E  =   \frac{1}{\alpha!} \Phi_{I}(\alpha) u^{\alpha}_{I},  \label{eq:expansion-multi-index}  
\end{eqnarray}
where $\alpha$ is  a {\bf cluster} comprised of $n$ lattice sites $\{  a_{1} \dots  a_{n}\}$, $I \equiv \{i_{1} \dots i_{n}\} $ are the corresponding Cartesian indices, and the summation over repeated indices is implied. The FCT $\Phi_{I}(\alpha)$ and displacement polynomial $u^{\alpha}_{I} \equiv  \prod_{k} u_{a_{k} i_{k}} $ are now referenced in this compact notation.  To avoid double counting, here the {\em order} in which to reference $\alpha$ has to be unambiguous, e.g. pre-determined by natural ordering of indices without loss of generality. We call $\alpha$ a {\bf proper} cluster if it contains no duplicate sites (as used in the cluster expansion model \cite{Sanchez1984PA334, *deFontaine199433}), or  {\bf improper} otherwise, e.g.\ $\{a, a\}$.

\subsection{Independent FCT parameters}

As the Taylor expansion coefficients of the crystal potential energy, force constants have to satisfy some physical constraints, namely derivative commutativity, space group symmetry,  and translational and rotational invariance \cite{Horton1974dynamical}.  The number of independent FCT matrix elements is reduced by these constraints, especially in solids with high symmetry.

\subsubsection{Commutativity of partial derivatives}
According to Schwarz's theorem, the FCTs as partial derivatives defined by Eq.~(\ref{eq:deritative})
are commutative with respect to the order of the partial derivatives if the PES is sufficiently smooth.
We do not have to worry about this constraint for a proper cluster $\alpha$, since the order of distinctive sites is pre-determined. However, if $\alpha$ is improper, there exists some non-trivial one-to-one indexing function $\pi$, i.e., $\{\pi(1), \dots, \pi(n)\}$ is a permutation of $\{1, \dots, n\}$ that maps $\alpha$ to itself:
$
\pi(\alpha) = \alpha.
$
We have
\begin{eqnarray}
\label{eq:improper-permutation}
\Phi_{\pi(I)} (\alpha) = \Phi_{I}(\alpha) \ \ \forall \ \ \pi(\alpha) = \alpha.
\end{eqnarray}
For instance, it is well known that an improper pair FCT satisfies $\Phi_{xy}(a,a)= \Phi_{yx}(a,a)$, i.e., it is a symmetric matrix. This has significant impact on the functional form of the long-range electrostatic contributions to the force constants, as discussed in Ref.~\onlinecite{CSLD-phonon}.

\subsubsection{Space group symmetry}

In a crystalline solid, the potential energy is invariant under the operations of the crystal space group $S$.  As a consequence, FCTs of cluster $\alpha$ and its mapping $\hat{s} \alpha$ under a  symmetry operator $\hat{s}$ are linearly related by a $3^{n} \times 3^{n}$ matrix $\Gamma$:
\begin{eqnarray}
\Phi_{I}(\hat{s} \alpha)& =& \Gamma_{IJ}(\hat{s})  \Phi_{J}(\alpha), \label{eq:FCT-symmetry}  \\
\Phi_{I}(\alpha) &=& \Gamma_{IJ}(\hat{s}^{-1})  \Phi_{J}(\hat{s}\alpha). 
\end{eqnarray}
To see this, first consider a proper cluster with a certain pre-determined ordering, $\alpha=\{ a_1 \dots  a_n \}$. In Cartesian coordinates, operation $\hat{s}$ consists of an orthogonal transformation by a $3 \times 3$ matrix $\gamma$, followed by a translation $\boldsymbol \tau$:
\begin{equation*}
{\bf r}' \equiv \hat{s} {\bf r} = {\mathbb \gamma} \cdot {\bf r} + {\boldsymbol \tau}.
\end{equation*}
It maps $\alpha$ one-to-one into $\{ \hat{s} a_{1} \dots \hat{s} a_{n} \}$, which  is in general {\em not} ordered, but rather a permutation of the ordered $\alpha' \equiv \hat{s} \alpha =\{a'_{1}, \dots, a'_{n}\}$, which can be referenced from the former by $a'_j = \hat{s} a_{\pi(j)}$, where $\pi$ is an indexing function.

For clarity, here we simplify the labels: ${\bf u}_j \equiv {\bf u}_{a_j}$. After $\hat s$, the displacement at the site $j$ of $\alpha'$ is
\begin{equation}
{\bf u}'_{j} = {\boldsymbol \gamma} \cdot {\bf u}_{\pi(j)}. \nonumber
\end{equation}
The potential energy of the original cluster is 
\begin{eqnarray}
\nonumber
& & {\boldsymbol \Phi}(\alpha) \cdot \left( {\bf u}_{1} \otimes {\bf u}_{2} \otimes \ldots \otimes {\bf u}_{n} \right) \\
\label{eq:Eorig}
&\equiv& \Phi_{i_1 \ldots i_n}(\alpha) u_{1, i_1}  u_{2, i_2} \ldots u_{n, i_n}. \nonumber
\end{eqnarray}
After transformation it becomes
\begin{eqnarray*}
\nonumber
& & {\boldsymbol \Phi} (\alpha') \cdot \left( {\bf u}'_{1} \otimes  {\bf u}'_{2} \otimes \ldots \otimes {\bf u}'_{n} \right) \\
\nonumber
& = & \Phi_{j_1 \ldots j_n} (\alpha')
	\gamma^{j_1}_{i_1} u_{\pi(1), i_1} \ldots \gamma^{j_n}_{i_n} u_{\pi(n), i_n}  \nonumber \\
 &= &\Phi_{j_1 \ldots j_n} (\alpha')
	\gamma^{j_1}_{i_{\pi(1)}} u_{\pi(1), i_{\pi(1)}} \ldots \gamma^{j_n}_{i_{\pi(n)}} u_{\pi(n), i_{\pi(n)}} \nonumber \\
&=&\Phi_{j_1 \ldots j_n} (\alpha')
	\gamma^{j_1}_{i_{\pi(1)}} \cdots \gamma^{j_n}_{i_{\pi(n)}}    u_{1, i_{1}} \cdots  u_{n, i_{n}},
\end{eqnarray*}
where we changed summation indices. 
Comparing the above expressions, the following must hold:
\begin{eqnarray}
\label{eq:PhiSymm1}
\Phi_{i_1\ldots i_n} (\alpha)&=& \gamma^{j_1}_{ i_{\pi(1)}}  \cdots  \gamma^{j_n}_{ i_{\pi(n)}} \Phi_{j_1 \ldots j_n} (\hat{s}\alpha) .
\end{eqnarray}
or
\begin{eqnarray}
\label{eq:GammaJI}
\Gamma_{IJ}(\hat{s}^{-1}) &=& \gamma^{j_1}_{i_{\pi(1)}}  \cdots  \gamma^{j_n}_{i_{\pi(n)}}
\end{eqnarray}
Since the matrix $\gamma$ of symmetry operation $\hat{s}$ is orthogonal,
\begin{eqnarray}
\label{eq:GammaIJ}
\Gamma_{IJ}(\hat{s}) &=& \bar{\gamma}^{j_1}_{i_{\pi^{-1}(1)}}  \cdots  \bar{\gamma}^{j_n}_{i_{\pi^{-1}(n)}}  = \gamma^{i_{\pi^{-1}(1)}}_{j_1}  \cdots \gamma^{i_{\pi^{-1}(1)}}_{j_1} \nonumber \\
&=& \gamma^{i_{1}}_{j_{\pi(1)}}  \cdots \gamma^{i_{n}}_{j_{\pi(n)}} = \Gamma_{JI}(\hat{s}^{-1}).
\end{eqnarray}
$\Gamma$ is therefore also orthogonal.

Additionally, if cluster $\alpha$ is improper, $\pi$ is not unique but rather allows arbitrary permutations of the indices belonging to any repeated site. 
Taking into account the commutativity relation in Eq.~(\ref{eq:improper-permutation}), the above derivation obviously holds for improper clusters.

In particular, the FCTs remain unchanged under translations by integer combinations of lattice vectors $\mathbf{b}_{k}$ ($\gamma=\mathbbm{1}$, $\Gamma = \mathbbm{1}$):
\begin{eqnarray}
 \Phi_{I}(\alpha +  N_{k} \mathbf{b}_{k})  \equiv  \Phi_{I}(\alpha +    \mathbf{R} ) = \Phi_{I}(\alpha) \label{eq:sgop-translation}.
\end{eqnarray}

We divide the set of all clusters of lattice sites into {\bf orbits} under the space group $S$. The orbit of cluster $\alpha$ is defined as the set of clusters into which $\alpha$ can be transformed by $S$:
\begin{eqnarray}
S \alpha \equiv \{ \hat{s} \alpha | \hat{s} \in S \}  \nonumber 
\end{eqnarray}
Given the linear relationship in Eq.~(\ref{eq:FCT-symmetry}) between symmetrically equivalent clusters, determination the FCTs of an entire orbit $S\alpha$ is reduced to (1) finding the FCT of a representative or symmetrically distinctive cluster $\alpha$, e.g. by selecting one with as many lattice sites in the primitive cell as possible,
and (2) calculating the FCTs of other clusters in $S \alpha$ by applying Eqs.~(\ref{eq:FCT-symmetry}) and (\ref{eq:GammaIJ}).  Hence it suffices to focus on the FCTs of all symmetrically distinctive representative clusters. The latter, essentially the set of all orbits under $S$ and hereafter used indistinguishably, will be called the {\bf orbit space} $A/S$ where $A$ denotes the set of all possible clusters. Now Eq.~(\ref{eq:expansion-multi-index}) can be rewritten as
\begin{eqnarray}
E&=& \sum_{ \alpha \in A/S} \frac{1}{\alpha!} \sum_{\alpha' \in S\alpha }   \Phi_{I}(\alpha'= \hat{s} \alpha) u^{\alpha'}_{I} \nonumber \\
& =&\sum_{ \alpha \in A/S} \frac{1}{\alpha!} \sum_{\hat{s} \alpha \in S\alpha }  \Gamma_{IJ}(\hat{s}) \Phi_{J}(\alpha) u^{\hat{s} \alpha}_{I},  \label{eq:potential-by-orbit} 
\end{eqnarray}
where the first summation is over all representative $\alpha \in A/S$ and the second one over clusters in the orbit $S \alpha$.

Given a representative cluster $\alpha$, the FCT may be further simplified. We define the {\bf isotropy group} $S_{\alpha}$ as the subset of $S$ that maps $\alpha$ to itself:
\begin{eqnarray*}
S_{\alpha} \equiv \{ \hat{s} \in S | \hat{s} \alpha = \alpha \}.   \label{eq:isotropy}  
\end{eqnarray*}
For instance, $S_{\alpha}$ of an improper site cluster $\alpha=\{a,\dots,a\}$ is identical to the point group of site $a$. According to Eq.~(\ref{eq:FCT-symmetry}), $\mathbf{\Phi}(\alpha)$ satisfies 
\begin{eqnarray}
\Phi_{I}(\alpha) &=& \Gamma_{IJ} (\hat{s}) \Phi_{J}(\alpha), \  \forall \hat{s}\in S_{\alpha}.
\label{eq:isotropy-constraint}  
\end{eqnarray} 

As a simple example, consider pair interactions in a periodic crystal with one lattice site per unit cell. Each site is at the center of inversion, and hence each pair is transformed onto itself upon inversion followed by a translation with ${\boldsymbol \tau} = -{\mathbf R}$, where $\mathbf R$ is the lattice vector pointing from the origin to the second vertex of the pair. Since such an operation permutes the vertices of the pair, application of Eq.~(\ref{eq:PhiSymm1}) results in the well-known symmetry condition for pair interactions in a monoatomic crystal:
\begin{equation}
\label{eq:Pairsymm}
{\boldsymbol \Phi}_\text{pair} = {\boldsymbol \Phi}^T_\text{pair}. \nonumber
\end{equation}
This also holds in multicomponent crystals for pair interactions between equivalent lattice sites which are at the center of inversion and separated by a lattice vector. 

Consider as another example the FCT $\boldsymbol{\Phi}(a,\dots,a)$ in diamond cubic silicon with point group $\bar{4}3m$ and rock-salt NaCl with point group $m \bar{3} m$. The non-zero elements are shown in Table~\ref{tab:FCT-Si-NaCl}. As expected, harmonic force constants $\boldsymbol{\Phi}(aa)$ are constrained by symmetry to be $\Phi_{xx}= \Phi_{yy} = \Phi_{zz}$
, i.e.\ isotropic. The anharmonic FCTs are more complicated. In contrast to Si, inversion symmetry in rock-salt eliminates the odd order FCT.
\begin{table}[htbp] 
	\begin{ruledtabular}
		\begin{tabular}{|p{0.3 \linewidth}|p{0.3 \linewidth}|p{0.3 \linewidth}|}
		\multicolumn{3}{|c|}{ Si} \\
		\hline
 1: 0  & 2: 1		&    3: 1 \\ 
None	& $\Phi_{xx}= \Phi_{yy} = \Phi_{zz}$  or $xx=yy=zz$ & $xyz= xzy= \dots= zyx$ \\
\hline
 4: 2 &  5: 1 &  6: 3 \\
$xxyy= xyxy = xyyx= \dots= zzyy$; $xxxx=yyyy=zzzz$ & $xxxyz= xxxzy =xxyxz= \dots= zzzyx$	& $xxxxxx= yyyyyy = zzzzzz$; $xxyyzz=\dots = zzyyxx$; $xxxxyy=\dots = zzzzyy$ \\
\hline
		\multicolumn{3}{|c|}{ NaCl} \\
		\hline
1: 0 & 2: 1		&    3: 0 \\ 
None	&  Same as Si &  None \\
\hline
 4: 2 &  5: 0 &  6: 3 \\
Same  &  None	& Same 
		\end{tabular}
	\end{ruledtabular}
	\caption{Non-zero elements of $n$-th order $\boldsymbol{\Phi}(a \dots a)$ in Si and NaCl after symmetrization by isotropy group. The number of degrees of freedom left is shown after $n$. The results for Na and Cl are identical since they have the same point group symmetry. Note here $\Phi_{xx}$ is written as $xx$ for brevity.
	\label{tab:FCT-Si-NaCl}} 
\end{table}

\subsubsection{Translational invariance}
According to the Noether's theorem, translational invariance of the Hamiltonian accounts for the conservation of linear momentum. Irrespective of the lattice type, the invariance of the total energy upon an arbitrary, uniform translation of the crystal leads to the acoustic sum rule (ASR) for the pair force constants:
\begin{eqnarray}
{\boldsymbol \Phi}(\{aa\}) = - \sum_{b \neq a} {\boldsymbol \Phi}(\{ab\}) \label{eq:ACsumrule}.
\end{eqnarray}
which states that the FCT of the improper pair cluster on lattice site $a$ can be obtained by summing up the FCTs of all proper pairs $\{a,b\}$.
The ASR can be generalized to any order as:
\begin{eqnarray}
\sum_{a} \Phi_{I}(\{a, b, c, \cdots \}) = 0  \nonumber \label{eq:translational-invariance}  
\end{eqnarray}
for arbitrary lattice sites $b, c, \cdots$ and cartesian indices $I$. 
Similar to Eq.~(\ref{eq:potential-by-orbit}), the above summation can be rewritten by grouping clusters into orbits
\begin{eqnarray}
 \sum_{a} \Phi_{I}(\hat{s} \alpha =\{a, b  \cdots \}) =  \sum_{a}  \Gamma_{IJ}(\hat{s})\Phi_{J} (\alpha)=0, \label{eq:translational-invariance-orbit}  
\end{eqnarray}
where any cluster $\{a, b  \cdots \}$ is identified as related to a representative cluster $\alpha \in A/S$ by the operator $\hat{s}$.
This constitutes yet another set of linear symmetry constraints on the FCTs.

In analogy to the above discussion, conservation of the angular momentum requires rotational invariance of the total Hamiltonian, which can also be expressed as linear constraints on the FCTs. However, rotational invariance involves force constants of different order in the same equation and is difficult to implement numerically. In this work, we do not impose rotational invariance constraints explicitly, relying instead on the fact that the DFT calculations themselves are rotationally invariant and should result in a Taylor expansion that obeys the rotational symmetry to a good accuracy.

\subsubsection{Determination of independent parameters}
As discussed previously, invariance with respect to translation by a lattice vector in Eq.~(\ref{eq:potential-by-orbit}) allows us to focus on the FCTs of representative clusters $\alpha \in A/S$. We denote by $\mathbf{\Phi}_{S}$ the one-dimensional combined list of all $N_{\Phi}$ such FCT elements. The number of truly independent parameters can be further reduced by taking into account the symmetries of cartesian indices for repeating vertices of improper clusters in Eq.~(\ref{eq:improper-permutation}), the space group symmetry constraints for the isotropy group $S_{\alpha}$ in Eq.~(\ref{eq:isotropy-constraint}), as well as the translational invariance constraints in Eq.~(\ref{eq:translational-invariance-orbit}). All three equations can be expressed in a linear equation for $\mathbf{\Phi}_{S}$:
\begin{eqnarray}
\mathbb{B} \mathbf{\Phi}_{S} =0,
   \label{eq:linear-constraint} 
\end{eqnarray}
where the matrix $\mathbb{B}$ contains the above mentioned (possibly redundant) constraints. For example, Eq.~(\ref{eq:isotropy-constraint}) can be rewritten as
$$
\left[ \Gamma(\hat{s}) - \mathbbm{1} \right] \boldsymbol{\Phi}(\alpha) =0.
$$
Note that Eqs.~(\ref{eq:improper-permutation}) and (\ref{eq:isotropy-constraint}) symmetrize a single FCT while Eq.~(\ref{eq:translational-invariance-orbit}), the translational invariance, relates different tensors of the same order. 

The basis vectors of the null-space of matrix $\mathbb{B}$ in Eq.~(\ref{eq:linear-constraint}) can be used to identify independent FCT parameters. Depending on the null-space construction method, the choice of independent parameters may not be unique. In this work we employ an iterative row-reduction algorithm with the three sets of constraints applied in the order given in the previous paragraph.  If the null-space dimension or {\bf nullity} of $\mathbb{B}$ is $N_{\phi}$, we are left with $N_{\phi}$ independent parameters $\boldsymbol{\phi}$ with which to express the original $N_{\Phi}$ variables:
\begin{eqnarray}
\mathbf{\Phi}_{S}  = \mathbb{C} \boldsymbol{\phi} ,
   \label{eq:independent-variables} 
\end{eqnarray}
where $\mathbb{C}$ is a $N_{\Phi} \times N_{\phi}$ matrix. More details on this procedure can be found in Appendix \ref{sec:null-space}. It allows us to impose the physical constraints exactly, without having to check the numerical accuracy of, e.g.,  the ASR. 

As an example, consider a minimal model of anharmonic lattice dynamics for silicon. It consists of nearest neighbor interactions with two symmetrically distinct pairs $\{aa\}$ and $\{ab\}$ and two anharmonic triplets $\{aaa\}$ and $\{aab\}$, where the lattice sites are $a=(000)$ and $b=\left(\frac14 \frac14 \frac14\right)$. There are only two independent harmonic parameters and three anharmonic ones:
\begin{enumerate}
\item $\mathbf{\Phi}(aa) = -4 \phi_1 \mathbbm{1}$ and $\mathbf{\Phi}(ab) = \phi_1 \mathbbm{1}$;
\item $\mathbf{\Phi}(ab) = \phi_2 \left( \begin{array}{ccc}
0 & 1 & 1 \\
1 & 0 & 1 \\
1 & 1 & 0 \end{array} \right)
$;
\item $\Phi_{ijk}(aaa) = -4 \phi_3 |\epsilon_{ijk}|$, $\Phi_{ijk}(aab) = \phi_3 |\epsilon_{ijk}|$;
\item $\Phi_{ijk}(aab) = \phi_4 \left(1- |\epsilon_{ijk}| - \delta_{ijk} \right)$;
\item $\Phi_{ijk}(aab) = \phi_5   \delta_{ijk}  $,
\end{enumerate}
where $\epsilon_{ijk}$ is the Levi-Civita function and $\phi_k$'s are the final independent parameters we seek. Obviously $\boldsymbol{\Phi}(a a)$ and $\boldsymbol{\Phi}(a a a)$ are consistent with Table \ref{tab:FCT-Si-NaCl}.

\subsection{Linear problem for force constants}
In order to calculate the FCTs, we take advantage of the force-displacement relationship. 
The force $F_{\mathbf{a}}$ on lattice site $a$ in direction $i$ can be obtained from taking the derivative of Eq.~(\ref{eq:potential-by-orbit}):
\begin{eqnarray}
F_{\mathbf{a}}&=& -\partial E / \partial u_{\mathbf{a}} \nonumber \\
&=& -\sum_{ \alpha \in A/S} \frac{1}{\alpha!} \sum_{\hat{s} \alpha \in S\alpha }  \Gamma_{IJ}(\hat{s}) \Phi_{J}(\alpha)  \partial_{\mathbf{a_{0}}} u^{\hat{s} \alpha}_{I} .
\label{eq:force-MI} 
\end{eqnarray}
The forces on the left hand side can be obtained from first-principles calculations according to the Hellmann-Feynman theorem using any general-purpose DFT code for a set of atomic configurations in a supercell, similar to the direct method for harmonic force constants \cite{Parlinski1997PRL4063}. One may extract $3N_{a}-3$ force components in  a supercell of $N_{a}$ lattice sites \footnote{The other three components are discarded since all forces sum to zero due to translational invariance.},
leaving us with the desired linear problem $\mathbf{F} = \mathbb{A'} \mathbf{\Phi}_{S}$ between force components and FCT parameters. 
The so-called {\bf sensing} (or correlation) matrix $\mathbb{A}'$ of dimension $N_{F} \times N_{\Phi}$ is calculated from atomic displacements according to 
\begin{eqnarray}
\mathbb{A}'(\mathbf{a}, \alpha I) &=&-\frac{1}{\alpha!} \sum_{\hat{s} \alpha \in S\alpha }  \Gamma_{JI}(\hat{s})   \partial_{\mathbf{a}} u^{\hat{s} \alpha}_{J}.
   \label{eq:sensing-matrix} 
\end{eqnarray}
Considering the independent parameters from Eq.~(\ref{eq:independent-variables}), the final linear problem to solve is
\begin{eqnarray}
\mathbf{F} 
 = \mathbb{A}' \mathbb{C} \boldsymbol{\phi} \equiv \mathbb{A} \boldsymbol{\phi} ,
   \label{eq:linear-problem-final} 
\end{eqnarray}
and the sensing matrix $\mathbb{A}$ for independent variables $\phi$ is $N_{F} \times N_{\phi}$ dimensional. Once the desired $\boldsymbol{\phi}$ is obtained, any FCT can be found using Eqs.~(\ref{eq:FCT-symmetry}) and (\ref{eq:independent-variables}).

Alternatively, the linear equation to fit the total energy is, according to Eq.~(\ref{eq:potential-by-orbit}),
\begin{eqnarray}
\mathbf{E} 
&=& \mathbb{A}^\text{E} \mathbb{C} \boldsymbol{\phi},  \nonumber \\
A^\text{E} (\{u\}, \alpha I)&=&  \frac{1}{\alpha!} \sum_{\hat{s} \alpha \in S\alpha }  \Gamma_{IJ}(\hat{s})  u^{\hat{s} \alpha}_{I}.  \label{eq:sensing-matrix-energy} 
\end{eqnarray}
This was done for CSLD phonon calculations in $\delta$-Pu \cite{Soderlind2015SR15958} when accurate forces were not available.

\subsection{Pairwise potential between bonded atoms}
\begin{figure}[htp]
\includegraphics[width = 0.75 \linewidth]{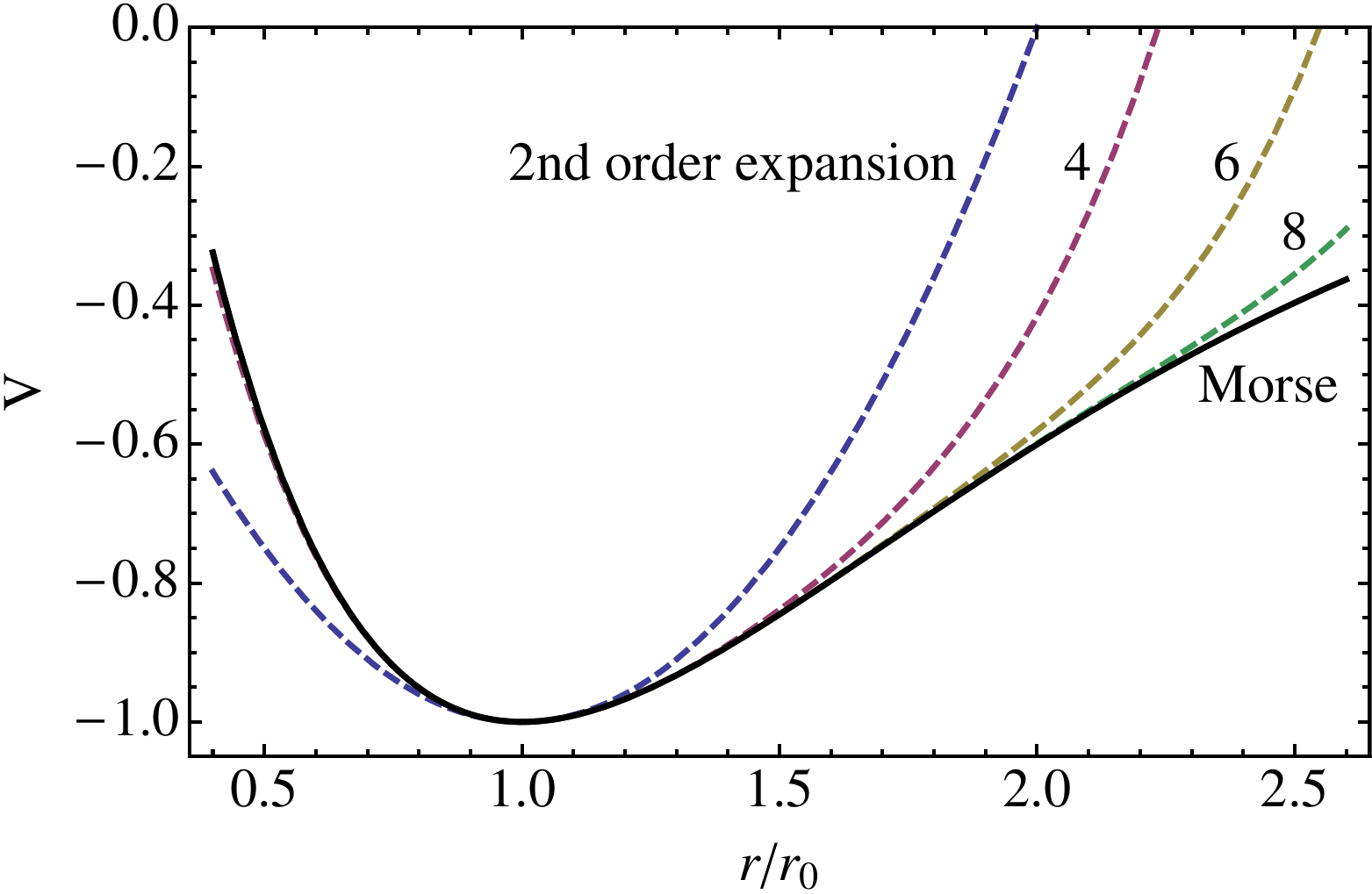}
\caption{Taylor expansion of the Morse potential, $V(r)= D_0 \left( e^{-2a(r-r_0)} - 2e^{-a(r-r_0)} \right) $ with $D_0=1$ and $a=1$.}
\label{fig:potential-taylor-expansion}
\end{figure}
The anharmonic force constants can be directly used to calculate phonon lifetimes and thermal properties (e.g., lattice thermal conductivity) in weakly anharmonic materials using the perturbation theory (PT) and the Boltzmann transport equation (BTE), as discussed in Sec.~\ref{sec:PT+BTE} below.  In cases of strong anharmonicity when the accuracy of PT+BTE starts to break down, one may wish to use multi-scale modeling techniques such as classical Monte Carlo (MC) or molecular dynamics (MD). In these cases, using a simple Taylor expansion of the PES as the Hamiltonian for the MC and MD simulations may lead to numerical problems with lattice stability. In this section, we discuss the cause of these stability issues and outline a practical solution.

First, the Taylor expansion is accurate only within a limited range of displacements, while in highly anharmonic crystals atoms commonly experience very large (a few 10\% of the NN distance) deviations from the equilibrium positions. This is seen by considering the one-dimensional Taylor expansion for the Morse potential that is sometimes used to model the interatomic interaction in a diatomic molecule; the results are shown  in Fig.~\ref{fig:potential-taylor-expansion}. At elevated temperature and large atomic deformations, there is substantial deviation in the 6th and even 8th order expansions, a reflection of the inherent limitation of the Taylor expansion. When the displacement is large enough with energy $\Delta E$, the expansion may completely break down, with probability proportional to $e^{- \Delta E/ k_B T}$. This probability increases quickly at high temperature, making a conventional Taylor expansion intrinsically problematic for multi-scale modeling.

Second, to improve the accuracy and applicable displacement range by increasing the maximum expansion order $n_{\max}$ is computationally demanding.  In spite of our numerically and algorithmically optimized implementation, we currently find it difficult to go beyond the 6th order. The next even order FCT contains $3^8=6561$ elements, and the transformation matrix $\Gamma $ is $3^8 \times 3^8$-dimensional, taking more than 300 Megabytes of memory as a double-precision matrix (though a sparse representation would cost substantially less). The sensing matrix $\mathbb{A}$, which is generally dense, can leave an even larger memory footprint. Since the number of clusters also explodes quickly with order $n$, one rapidly runs out of computer memory. In this work, we have chosen to truncate all Taylor expansions at $n_{\max}=6$ to keep the computations manageable, even though our formalism and implementation allow for higher orders. This choice is found to be more than sufficient for the weakly anharmonic systems like Si and NaCl at temperatures up to 600 K. In strongly anharmonic materials such as the thermoelectric compound Cu$_{12}$Sb$_4$S$_{13}$ (tetrahedrite) \cite{Zhou2014PRL185501}, this truncation may have a noticeable detrimental effect on the accuracy of the lattice dynamical model. Indeed, our MD simulations based on the 6th order expansion for tetrahedrite were susceptible to divergences due to unphysical covalent bond breaking at 300 K \cite{Zhou2014PRL185501}.

To solve this problem, we introduced pairwise force field (FF) potentials \cite{Zhou2014PRL185501} to augment the lattice dynamical model of tetrahedrite:
\begin{eqnarray}
E = E_{\mathrm{LD}} + E_{\mathrm{FF}} =  E_{\mathrm{LD}} + \sum_{a \leftrightarrow b} E_{ab}(r_{ab}),
\label{eq:LD+FF}
\end{eqnarray}
where $E_{\mathrm{LD}}$ is the normal Taylor expansion of Eq.~(\ref{eq:Taylor}),  the summation goes over covalently bonded sites $a, b$, and $r_{ab}= |\mathbf{r}_{a}- \mathbf{r}_{b}|$. Each pair potential $E_{ab}$ is expanded as
\begin{eqnarray}
  E_{ab}(r_{ab}) = \epsilon_{l} p_{l}((r_{ab} - r^0_{ab})/r_\text{c}),
\label{eq:radial-FF}
\end{eqnarray}
where $p_{l}$ and $\epsilon_l$ are the $l$-th single-variable basis function and the corresponding coefficient, respectively, and $r^0$ is the equilibrium bond length. In this work Legendre polynomials are used as basis functions. The correlation matrix $\mathbb{A}_{\mathrm{FF}}$ between force components and coefficients $\{\epsilon \}$ can be written for each  training structure 
\begin{eqnarray}
\mathbb{A}_{\mathrm{FF}}(\mathbf{a}, l) = -  p'_l ((r_{ab} - r^0_{ab})/r_\text{c}) \mathbf{r}_{ab}/(r_{ab} r_\text{c}),
\label{eq:sensing-FF}
\end{eqnarray}
and appended  to the LD correlation matrix in Eq.~(\ref{eq:linear-problem-final}) to fit the unknown parameters $\{ \phi \}$ and $\{\epsilon\}$ in an expanded linear equation: 
$$
\mathbf{F}= \left(\mathbb{A}, \mathbb{A}_{\mathrm{FF}} \begin{array}{ll}   \end{array} \right )   \left(\begin{array}{l} {\boldsymbol \phi} \\ {\boldsymbol \epsilon}   \end{array} \right ) .
$$ 
The main advantage of this optional step is that we gain some knowledge of high-order anharmonicity by adding only a few coefficients $\epsilon_l$ rather than $3^n$ FCT elements. In MC or MD simulations, the pair potential is continuously extrapolated outside a reasonable range with a functional form $E_{ab}(r) \sim 1/r^m$ where $m=1$ for bond stretching and $m=6$ for compression:
$$
    E_{ab}(r)= 
\begin{cases}
    \epsilon_{l} p_{l}((r - r^0)/r_\text{c}) ,& \text{if }  -r_{\mathrm{c}} \leq r-r^{0} \leq  r_{\mathrm{c}} \\
   e_{>} + x_{>}/r ,& \text{if }   r-r^{0} >  r_{\mathrm{c}} \\
   e_{<} + x_{<}/r^{6} ,& \text{if }   r-r^{0} < -  r_{\mathrm{c}} 
\end{cases},
$$
where for continuity $e_{<(>)}= \epsilon_{l} p_{l}(\mp 1) - x_{<(>)}/(r_0 \mp r_c)^m $, and $x_{<(>)}=-\epsilon_{l} p'_{l}(\mp 1) (r_0 \mp r_c)^{m+1}/m r_c$.

There appears to be a drawback with this hybrid LD-FF approach: since the Taylor expansion is based on a complete basis set, inclusion of another set of basis functions might be counter-productive because the obtained $\boldsymbol \phi$ and $\boldsymbol \epsilon$ coefficients are no longer uniquely determined. Indeed, the combined series of Eq.~(\ref{eq:LD+FF}) is not a basis, but an over-complete {\bf frame}. Fortunately, a complete basis is not a requisite condition for CS to work: tight frames are known to be compatible and are widely used in image processing\cite{Candes2004CPAM219}. Our results show that this combined LD-FF expansion can sustain accurate and robust MD simulation of Cu$_{12}$Sb$_4$S$_{13}$ for very long durations.

\subsection{Lattice Molecular dynamics} 
A classical MD program (LMD) with Eq.~(\ref{eq:Taylor}) or optionally Eq.~(\ref{eq:LD+FF}) as the interatomic potential has been developed.  Multiple methods were implemented for calculating $\kappa_L$, including the Green-Kubo linear response formula \cite{Green1954JCP398,Kubo1957JPSJ570}, reverse non-equilibrium MD (RNEMD) \cite{Muller-Plathe1997JCP6082} and homogenous non-equilibrium MD (HNEMD) proposed by Evans \cite{Evans1982PLA457}. While all methods yielded similar results, we found after extensive testing that HNEMD was the most efficient. In HNEMD, the equations of motion are modified so that the force on lattice site $a$ is given by
\begin{equation}
    \label{eq:eq-motion}
    \mathbf{F}_a= F_a - \sum\limits_{b} \mathbf{F}_{ab} \left( \mathbf{r}_{ab} \cdot \mathbf{F}_e \right) + \frac{1}{N} \sum\limits_{b,c} \mathbf{F}_{bc}  \left( \mathbf{r}_{bc} \cdot \mathbf{F}_e \right),
\end{equation}
where $F_a$ is the unmodified force calculated from Eq.~(\ref{eq:force-MI}) and $\mathbf{F}_{ab}$ is the force on site $a$ due to  $b$. Contributions from third- and higher-order interactions to $\mathbf{F}_{ab}$ were obtained by partitioning the energy evenly among all lattice sites in the cluster, including repeated sites. The external field $ \mathbf{F}_e$ has the effect of driving higher energy (hotter) particles with the field and lower energy (colder) particles against the field, while a Gaussian thermostat is used to remove the heat generated by $ \mathbf{F}_e$. This results in a non-zero average heat flux given by
\begin{equation}
  \left<\mathbf{J}(t)\right> =  -\beta V \int\limits_0^t ds \left< \mathbf{J}(t-s)  \otimes  \mathbf{J}(0)  \right> \cdot \mathbf{F}_e.
\end{equation}
As $\mathbf{F}_e \rightarrow 0$, one recovers the linear response limit described by the Green-Kubo formula \cite{Green1954JCP398,Kubo1957JPSJ570}. For cubic systems the external field can be set to $\mathbf{F}_e  = (0,0,F_z)$, and in the limit of $t \rightarrow \infty$ we get the following relation:
\begin{equation}
  \kappa_L = \frac{V}{k_B T^2} \int\limits_0^\infty dt \left< J_z(t) J_z(0) \right> = \lim\limits_{F_z \rightarrow 0} \frac{- \left< J_z(\infty) \right>}{TF_z}.
\end{equation}
The process then involves a series of simulations at varying external fields $\mathbf{F}_e$ and constant ${T}$, with a simple linear extrapolation to zero field resulting in the true $\kappa_L$.

\subsection{Perturbation theory for anharmonic lattice dynamics}
\label{sec:PT+BTE}

In contrast to the classical molecular dynamics in real space, perturbation theory for phonon-phonon interactions is formulated in reciprocal (momentum) space\cite{peierls1996quantum, Ziman1960electrons}. In the framework of the relaxation time approximation (RTA) \cite{Ziman1960electrons} to the Boltzmann transport equation, lattice thermal conductivity is obtained by summing contributions from all phonon modes:
	\begin{equation}\label{eq:kappa}
	\kappa_{L}^{ij}=\frac{1}{k_{B}T^2\Omega N} \sum_{\lambda} f^{0}_{\lambda} (f^{0}_{\lambda}+1) (\hbar \omega_{\lambda})^2 v_{\lambda}^{i} v_{\lambda}^{j} \tau_{\lambda},
	\end{equation}
where $N$ is the number of included phonon modes, $\Omega$ is the volume of the primitive cell, $f^{0}_{\lambda}$  is the Bose-Einstein distribution function, and $\omega_{\lambda}$, $v_{\lambda}^{i}$ and $\tau_{\lambda}$ are the frequency, group velocity and relaxation time of the phonon mode $\lambda$. The relaxation times are the reciprocals of the total scattering rates, which are calculated via the Fermi's golden rule. Considering only the intrinsic three-phonon scattering processes ($\lambda \pm \lambda^{\prime} \to \lambda^{\prime\prime}$), the single mode relaxation time can be expressed as \cite{Broido2007APL231922,ward,shengbte}
	\begin{equation}
	\frac{1}{\tau_{\lambda}^{0}}=\sum_{\lambda^{'}\lambda^{\prime\prime}}^{+}\Gamma_{\lambda\lambda^{\prime}\lambda^{\prime\prime}}^{+}+
	\frac{1}{2}\sum_{\lambda^{\prime}\lambda^{\prime\prime}}^{-}\Gamma_{\lambda\lambda^{\prime}\lambda^{\prime\prime}}^{-}
	\end{equation}
where $\Gamma_{\lambda\lambda^{\prime}\lambda^{\prime\prime}}^{+}$ and $\Gamma_{\lambda\lambda^{\prime}\lambda^{\prime\prime}}^{-}$ are the scattering rates from the absorption ($+$) and emission ($-$) processes, which can be evaluated from the harmonic phonon dispersion and 3rd-order FTCs as follows:\cite{Broido2007APL231922,ward,shengbte}
	\begin{equation}
	\begin{split}
	\Gamma_{\lambda\lambda^{'}\lambda^{''}}^{\pm}=&\frac{\hbar\pi}{4} \left\{\begin{array}{c}  f_{\lambda^{'}}^{0}-f_{\lambda^{''}}^{0} \\ f_{\lambda^{'}}             ^{0}+f_{\lambda^{''}}^{0}+1  \end{array}\right\} \times \left| V_{\pm\lambda\lambda^{'}\lambda^{''}}^{(3)} \right|^{2} \\
	& \times \frac{\delta\left( \omega_{\lambda}\pm\omega_{\lambda^{'}}-\omega_{\lambda^{''}}  \right)}{\omega_{\lambda}\omega_{\lambda^{'}}\omega_{\lambda^{''}}}, \\
	\end{split}
	\end{equation}
	\begin{equation}
	\begin{split}
	V_{\pm\lambda\lambda^{'}\lambda^{''}}^{(3)}=&\sum_{a, \ell' b, \ell^{\prime\prime } c} \sum_{ijk}\Phi_{ijk}(0 a, \ell' b, \ell^{\prime\prime } c)\frac{\epsilon_{a,i}^{\lambda}\epsilon_{b,j}^{\pm\lambda^{\prime}} \epsilon_{c,k}^{-\lambda^{\prime\prime}}} {\sqrt{m_{a} m_{b} m_{c}}} \\
	 & \cdot e^{\pm i \mathbf{q}^{\prime} \cdot \mathbf{R}_{\ell'}} e^{-i\mathbf{q}^{\prime\prime}\cdot \mathbf{R}_{\ell^{\prime \prime}}},
	\end{split}
	\end{equation}
where $\epsilon^{\lambda}_{a,i}$ is the $i$-th component of the polarization vector of lattice site $a$, and $\mathbf{q}$ represents the phonon wave vector of mode $\lambda$. Refined iterative scheme can be used to obtain more accurate relaxation times by taking into account the nonequilibrium states of interacting phonons,\cite{omini1,omini2,Broido2007APL231922,Li2014CPC1747}
	\begin{equation}
	\tau_{\lambda}=\tau_{\lambda}^{0} \left ( 1+\Delta_{\lambda} \right),
	\end{equation}
where $\Delta_{\lambda}$ is the deviation from the single mode relaxation approximation:
	\begin{equation}
	\begin{split}
	\Delta_{\lambda}  =& \frac{1}{N} \sum_{\lambda^{\prime}\lambda^{\prime\prime}}^{+} \Gamma_{\lambda\lambda^{'}\lambda^{''}}^{+}
	\left( \xi_{\lambda\lambda^{''}} \tau_{\lambda^{''}} - \xi_{\lambda\lambda^{'}} \tau_{\lambda^{'}}\right) \\
	& + \frac{1}{N} \sum_{\lambda^{'}\lambda^{''}}^{-} \frac{1}{2} \Gamma_{\lambda\lambda^{'}\lambda^{''}}^{-}
	\left( \xi_{\lambda\lambda^{''}} \tau_{\lambda^{''}} + \xi_{\lambda\lambda^{'}} \tau_{\lambda^{'}}\right).
	\end{split}	
	\label{eq:Delta}
	\end{equation}
Here, $N$ is the number of sampling points and $\xi_{\lambda\lambda^{'}} \equiv \omega_{\lambda^{'}} v_{\lambda^{'}}^{z} / \omega_{\lambda} v_{\lambda}^{z}$. An improved version of the above formulation has been developed recently  in Refs.~\onlinecite{Tianli2016,Tianli2017} by incorporating four-phonon scattering processes.

\section{Theory: Compressive sensing} \label{sec:CS}

In this section we focus on the numerical solution of the linear problem $\mathbf{F} = \mathbb{A} \boldsymbol{\phi}$ of Eq.~(\ref{eq:linear-problem-final}). $\mathbb{A}$ is an $N_{F} \times N_{\phi}$ matrix, where $N_{F}$ is the number of calculated force components, and $N_{\phi}$ is the total number of unknown model parameters. In practice, the latter may far exceed $N_{F}$, making Eq.~(\ref{eq:linear-problem-final}) underdetermined. A reasonable approach would be to choose $\boldsymbol{\phi}$ so that it reproduces the training data $\mathbf{F}$ to a given accuracy with the smallest number of nonzero FCT components, i.e., by minimizing the so-called $\ell_0$ norm $\| \boldsymbol{\phi} \|_0 \equiv \sum_{I, \phi_{I} \neq 0} 1$. Unfortunately, this is a computationally intractable combinatorial problem.

We have recently shown that a similar problem in alloy theory, the cluster expansion (CE) method for configurational energetics \cite{Sanchez1984PA334, deFontaine199433}, can be solved efficiently and accurately using compressive sensing \cite{Nelson2013PRB035125,Nelson2013PRB155105}. CS has revolutionized information science by providing a mathematically rigorous recipe for reconstructing $S$-sparse models (i.e., models with $S$ nonzero coefficients out of a large pool of possibles, $N$, when $S \ll N$) from only $O(S)$ number of data points \cite{Candes2005IEEETIT4203,Candes2006IEEETIT489,Candes2006CPAM1207}. Given training data, CS automatically picks out the relevant  signals (i.e., expansion coefficients in physics models) and determines their values {\it in one shot.\/} 
The linear problem $\mathbf{F} = \mathbb{A} \boldsymbol{\phi}$ in Eq.~(\ref{eq:linear-problem-final}) is solved by minimizing the $\ell_1$ norm of the coefficients, 
\begin{equation}
\label{eq:l1norm}
\| \boldsymbol{\phi} \|_1 \equiv \sum_{i} | \phi_{i} |,  \nonumber
\end{equation}
while requiring a certain level of accuracy for reproducing the data. The $\ell_1$ norm serves as a computationally feasible {\it continuous\/} approximation to the discrete $\ell_0$ norm and results in a tractable {\it convex\/} optimization problem with a guarantee of finding the global minimum.  Mathematically, the solution is found as
\begin{equation}
\label{eq:CSfit}
\boldsymbol{\phi}^\text{CS} = \arg \min_{\boldsymbol{\phi}} \| \boldsymbol{\phi} \|_1 + \frac{\mu}{2} \| \mathbf{F} - \mathbb{A} \boldsymbol{\phi} \|^2_2,
\end{equation}
where the second term is the usual sum-of-squares $\ell_2$ norm of the fitting error for the training data (in this case, DFT forces). The $\ell_1$ term drives the model towards solutions with a small number of nonzero FCT elements, and the parameter $\mu$ is used to adjust the relative weights of the $\ell_1$ and $\ell_2$ terms. Higher values of $\mu$ will produce a  least-squares like fitting at the expense of denser FCTs that are prone to over-fitting, while small $\mu$ will produce very sparse under-fitted FCTs, simultaneously degrading the quality of the fit. The optimal value of $\mu$ that produces a model with the highest predictive accuracy lies between the aforementioned extremes and can be determined by monitoring the predictive error for a leave-out subset of the training data which is not used in Eq.~(\ref{eq:CSfit}) \cite{Nelson2013PRB035125}. The predictive accuracy of the resulting model is then validated on a third, distinct set of DFT data, which we refer to as the ``prediction set''.  This procedure was described in detail in Ref.~\onlinecite{Nelson2013PRB035125}.

The main advantages of CLSD over other methods for model building are that it does not require prior physical intuition to pick out potentially relevant FCTs and the fitting procedure is stable with respect to both random and systematic noise \cite{Candes2006CPAM1207}.

To solve Eq.~(\ref{eq:linear-problem-final}) with CS, all the FCTs need to have the same unit of force. We use dimensionless displacements by substituting $u \rightarrow u/u_{0}$, where $u_{0}$ is conceptually a ``maximum'' displacement chosen to be on the order of the amplitude of thermal vibrations. An order-$n$ FCT is then scaled by $\Phi \rightarrow \Phi u_{0}^{n-1}$ in Eq.~(\ref{eq:linear-problem-final}).

\subsection{CS Solver}
\label{sec:CS-solver}
Development of solvers for sparse signal recovery remains an active area of research\footnote{For a expanding list of solvers, see \url{https://sites.google.com/site/igorcarron2/cs\#reconstruction}}.
We adopted the split Bregman algorithm \cite{Goldstein2009SIAMJIS323}, which was previously used in the CS fitting of the configurational cluster expansion model \cite{Nelson2013PRB035125}.
The convergence rate of the convex minimization step in the split Bregman algorithm is strongly influenced by the condition number of $\mathbb{Q} = \mathbb{A}^T \mathbb{A} + \lambda \mu \mathbb{I}$.
It can be improved significantly by using a suitable preconditioner.

{\it Right preconditioner:\/} If one computes $s$ largest eigenvectors of $\mathbb{A}^T \mathbb{A}$, an efficient preconditioner can be constructed using 
\begin{equation}
\label{eq:rightprecond}
\mathbb{C}_p = \mathbb{V}^T (\mathbb{D} + \lambda \mu \mathbb{I})^{-\frac{1}{2}} \mathbb{V} + 
	(\lambda \mu )^{-\frac{1}{2}} \mathbb{N}^T \mathbb{N},
\end{equation}
where $\mathbb{D}$ is an $s \times s$ diagonal matrix of eigenvalues, $\mathbb{V}$ is an $s \times N_\phi$ matrix of the corresponding eigenvectors, and $\mathbb{N}$ is an $(N_\phi-s) \times N_\phi$ matrix containing the vector space complement of $\mathbb{V}$. In our experience, appreciable speed-up can be achieved even if $s$ is a fraction of the number of equations $N_F$ (typically, 1/4) due to rescaling of the few largest eigenvalues of $\mathbb{A}^T \mathbb{A}$. After variable substitution $\boldsymbol \phi = \mathbb{C}_p \boldsymbol \phi' $, the problem to be solved becomes $\mathbf{F}= \mathbb{A} \mathbb{C}_p \boldsymbol \phi'$. The main advantage of the right preconditioner is that the objective function remains the original sum-of-squares of residuals, and the expense of the preconditioning step versus the split Bregman iteration can be controlled by selecting $s$, the number of eigenvalues to be computed to construct the preconditioner.

\subsection{Training structures}
\label{sec:structures}
A key ingredient of CSLD is the choice of atomic configurations for the training and prediction sets. One of the most profound results of CS is that a near-optimal signal recovery can be realized by using sensing matrices $\mathbb{A}$ with {\it random\/} entries that are independent and identically distributed (i.i.d.) \cite{Candes2008IEEESPM21}.   For the discrete orthogonal basis in the CS cluster expansion \cite{Nelson2013PRB035125,Nelson2013PRB155105}, i.i.d. sensing matrices $\mathbb{A}$ could be obtained by enumerating all ordered structures up to a certain size and choosing those with correlations that map most closely onto quasi-random vectors on the unit sphere.  
Since the Taylor expansion employs non-orthogonal and unnormalized basis functions of a continuous variable, $u^{n}$, this strategy is difficult to adapt for CSLD and it becomes challenging to construct training configurations that give $\mathbb{A}$ with quasi-i.i.d. entries. As a result, we require larger training sets. Nevertheless, this is not a serious limitation because a large number of independent forces ($3m-3$) can be extracted from each $m$-atom supercell configuration.

It is intuitively appealing to use snapshots from {\it ab initio}  MD (AIMD) trajectories since they represent low-energy configurations. However, physically accessible low-energy configurations under thermodynamic distributions  give rise to strong cross-correlations between the columns of $\mathbb{A}$ (i.e., high mutual coherence of the sensing matrix \cite{Donoho2001IEEETIT2845}), which decreases the efficiency of CS due to the difficulty of separating correlated contributions to $\mathbf{F}$ from different FCTs.  To solve this conundrum, we combine the physical relevance of AIMD trajectories with the mathematical advantages of efficient compressed sampling by adding random displacements ($\sim$ 0.1--0.2 \AA) to each atom in sufficiently spaced snapshots from short AIMD trajectories. This procedure was found to decrease the coherence as measured by the cross-correlations between the columns of the sensing matrix $\mathbb{A}$ \cite{Donoho2001IEEETIT2845}, resulting in stable signal recovery.

For the relatively simple task of fitting harmonic force constants for phonon spectra, we found it sufficient to independently displace all atoms in the training supercell structure in random directions  by 0.01 \AA\ away from their equilibrium positions.

\subsection{Fitting in steps} \label{sec:fit-in-steps}
In practical CSLD model building, one may wish to adopt a divide-and-conquer strategy to fit different groups of parameters in steps. There are two reasons for this. First, as a consequence of the non-i.i.d. character of the sensing matrix $\mathbb{A}$, the numerical stability of fitting is somewhat reduced. Second, the number of independent parameters $N_\phi$ can easily exceed $10^4$ in complex structures with high-order anharmonicity, making direct fitting in one shot very inefficient. If $N_\phi$ is large ($\gtrsim$3000), one may therefore divide $\boldsymbol \phi$ into subsets $\boldsymbol \psi _1, \boldsymbol \psi _2, \dots$ and the sensing matrix  into corresponding sub-matrices $\mathbb{A}= \left(\mathbb{A}_1, \mathbb{A}_2, \dots \right)$, e.g.\  pairwise potential parameters $\boldsymbol \epsilon$, harmonic parameters $\boldsymbol \phi^{(2)}$, third-order $\boldsymbol \phi^{(3)}$, etc. The parameters $\boldsymbol \psi_n$ are fitted sequentially, taking into account the contributions of the previous ones:
\begin{align}
\mathbf{F}- \left(\mathbb{A}_1, \dots, \mathbb{A}_{n-1} \right) \left(\boldsymbol \psi _1^T, \dots, \boldsymbol \psi _{n-1}^T\right)^T= \mathbb{A}_n  \boldsymbol \psi _n.
\end{align}
Training structures can be adapted for each set of parameters: small displacements for the harmonic parameters $\boldsymbol \phi^{(2)}$, and gradually larger displacements from higher temperature AIMD snapshots for the higher order anharmonic terms. The is procedure was used for fitting \ce{Cu12Sb4S13}, as detailed in Section \ref{sec:results-test}.

\section{Results and discussions} \label{sec:results}

All DFT calculations were performed using the Perdew-Becke-Ernzerhof (PBE) functional \cite{Perdew1996PRL3865}, PAW potentials \cite{Blochl1994PRB17953}, a cutoff energy of 600 eV, energy convergence tolerance of $10^{-9}$ eV per atom, and no symmetry constraints as implemented in the VASP code \cite{Kresse1999PRB1758}. Lattice parameters were fixed at experimental values except for silicon clathrates (\ce{Si46}, \ce{Na8Si46}, and \ce{Ba8Si46}), which were fully relaxed. AIMD simulation was run in 1 fs steps with lowered computational accuracy (smaller cutoff and larger tolerance) and snapshots were taken at 3ps intervals and re-calculated with high accuracy.

\subsection{Tests on Si, NaCl, Al, Cu$_{12}$Sb$_4$S$_{13}$} \label{sec:results-test}

\begin{figure}[htp]
	\includegraphics[width = 0.85 \linewidth]{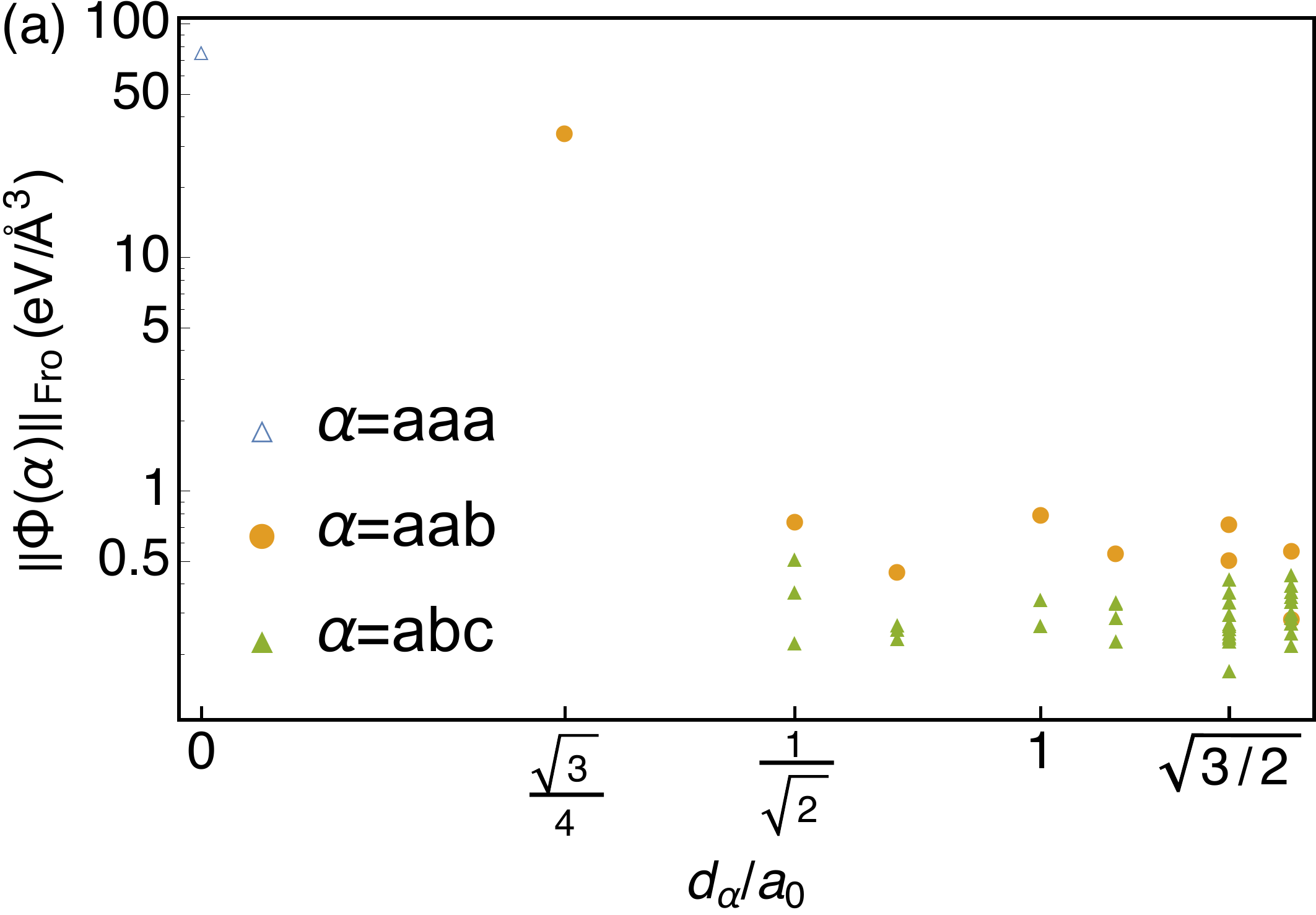}
	\includegraphics[width = 0.85 \linewidth]{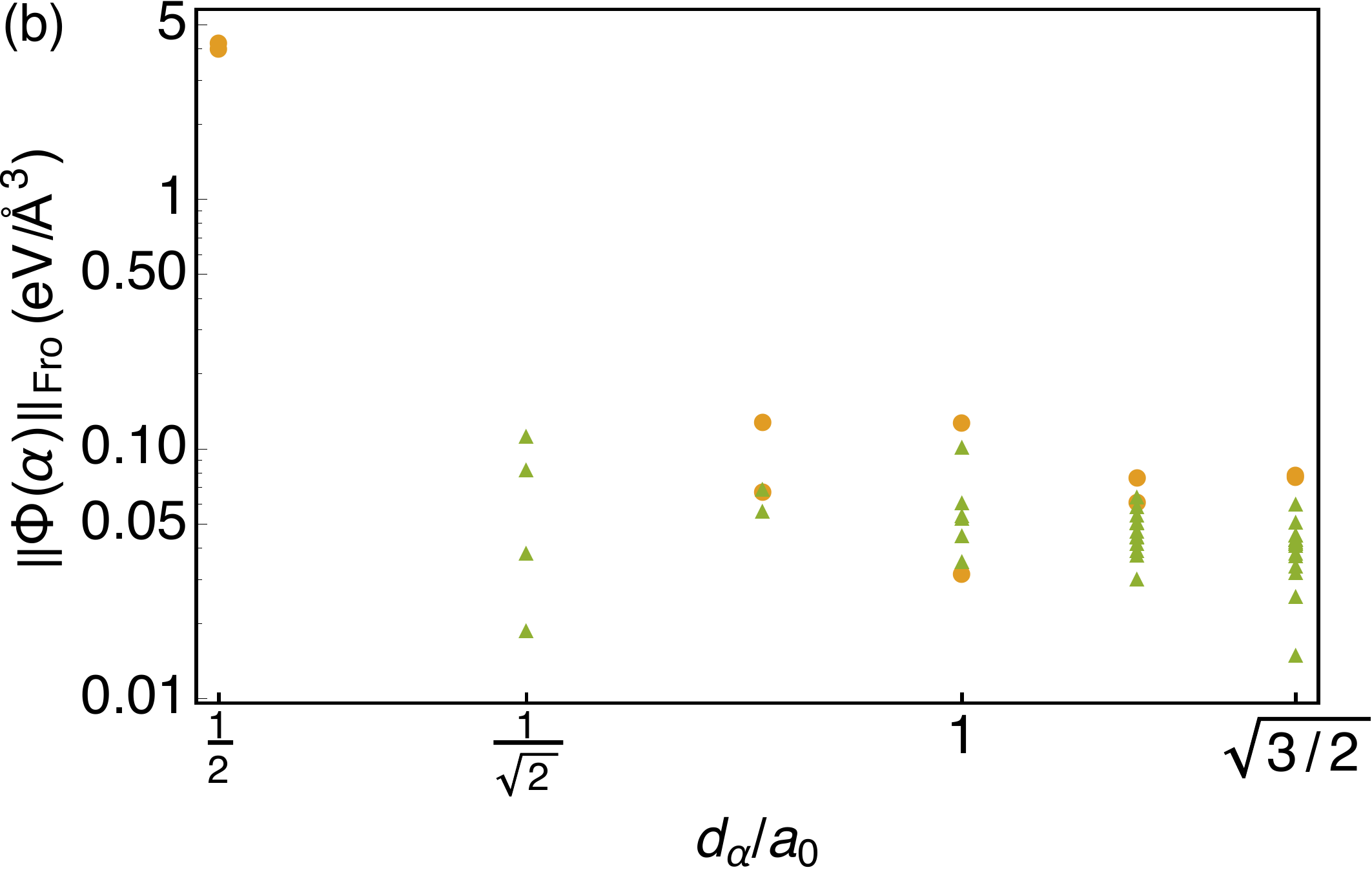}
	\includegraphics[width = 0.85 \linewidth]{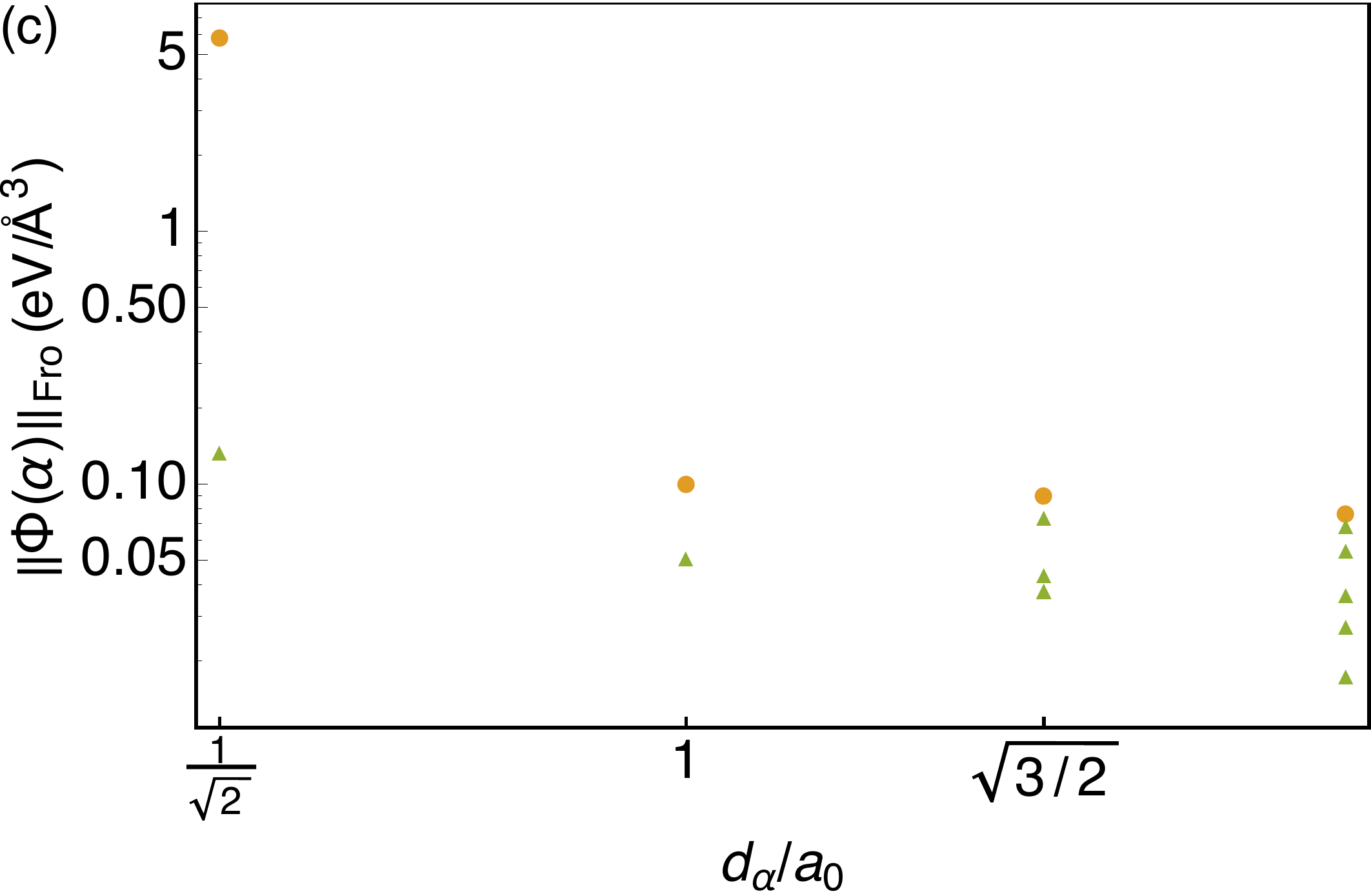}
	\caption{Log-linear plots of the Frobenius norm of the 3rd order FCT $\mathbf{\Phi}(\alpha)$ {\it vs.\/} the scaled distance $d_\alpha$ for (a) Si, (b) NaCl, and (c) Al.}
	\label{fig:third-order}
\end{figure}
We show in Fig.~\ref{fig:third-order} the results of third order fitting for cubic Si, rock salt NaCl, and fcc Al. In each case two $3\times 3 \times 3$ fcc supercell configurations with all atoms randomly displaced by 0.03 \AA\ were used for CSLD fitting. The fit included all pair and third-order clusters with diameter not exceeding half the size of the cell (here the diameter is defined as the maximum distance of pairs in the cluster, $d_{\alpha}=\max_{a,b \in \alpha}d_{ab}$).  In silicon, by far the most significant third order contribution is the improper on-site cluster $aaa$, which does not vanish due to the absence of inversion symmetry on the Si site. We find that the magnitudes of the FCTs involving two atoms ($aab$, filled circles) are generally larger than those of the proper clusters $abc$ (filled triangles). This feature is also observed in the harmonic force constants and can be attributed to the generalized ASR in Eq.~(\ref{eq:translational-invariance}), which relates the improper FCTs to proper and ``less improper'' ones.   $\mathbf{\Phi}(aab)$ drops in magnitude quickly as the interaction distance increases (note the log scale). In fact, the most appreciable $\mathbf{\Phi}(aab)$ on the nearest-neighbor $ab$ pair is 30--60 times larger than the second nearest neighbor and beyond, suggesting that the anharmonic FTCs are much more short-ranged than the harmonic ones.

\begin{figure}[htp]
	\includegraphics[width = 0.98 \linewidth]{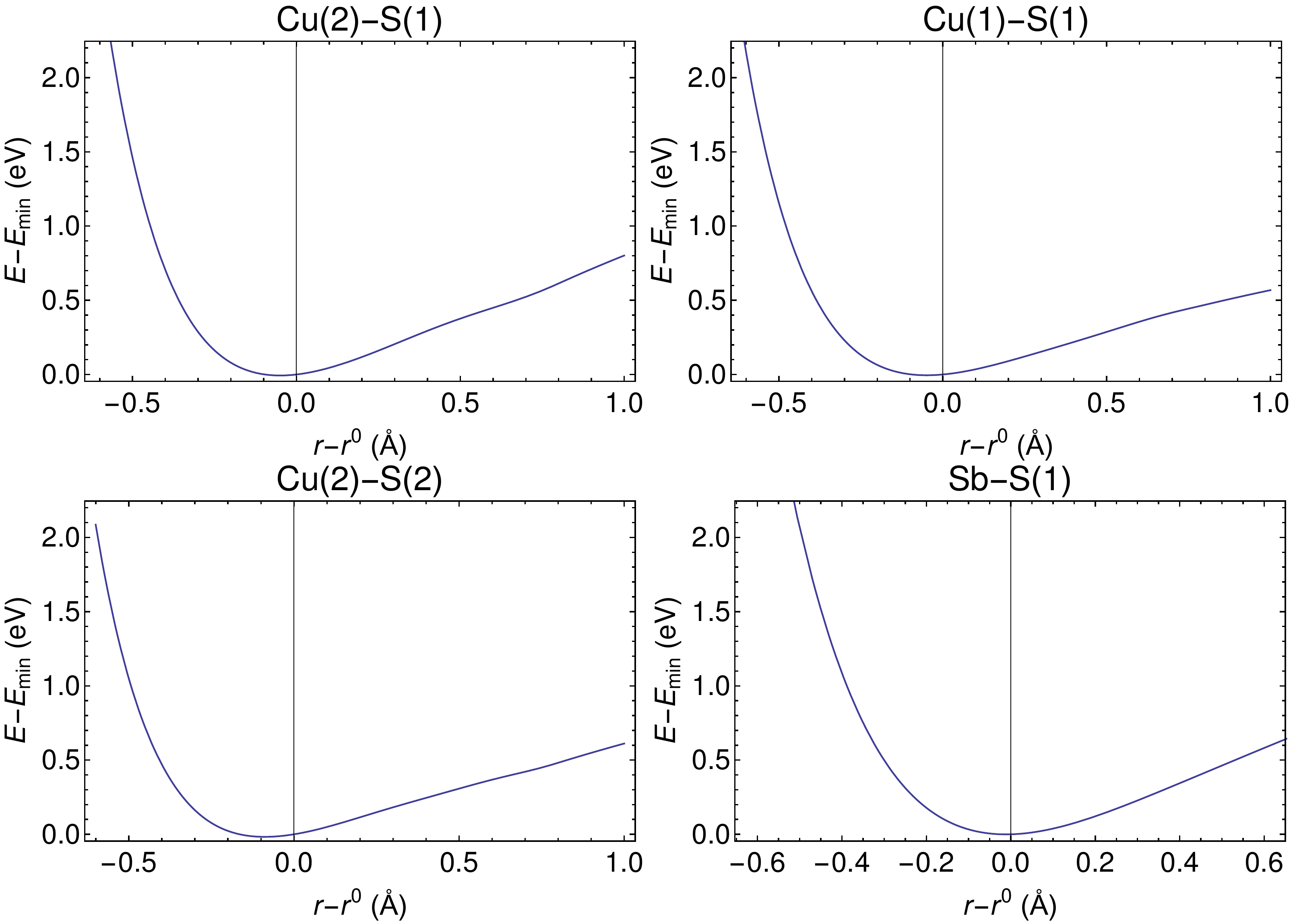}
	\caption{Fitted pairwise potential $E_\text{FF}$ for \ce{Cu12Sb4S13}.}
	\label{fig:tetrahedrite-pairpotential}
\end{figure}
The earth-abundant natural mineral tetrahedrite (\ce{Cu12Sb4S13}) has been shown to be a high-performance thermoelectric due to its very low lattice thermal conductivity and high thermoelectric power factor \cite{Lu2013AEM342, Suekuni2012APE051201}. The body-centered cubic (bcc) structure with space group  $I\bar{4}3m$ has 29 atoms in the primitive cell, a large number that complicates the computation of FCTs. For example, there are 188 distinct atomic pairs  within a radius of $a=10.4$~{\AA}, 116 triplets within $a/2$, etc. Taking into account the $3^{n}$ elements of each tensor, the number of unknown FCT coefficients is very large (55584 in our setting). After symmetrization, this is reduced to $3188$, which still represents a formidable numerical challenge. 
The fitting for Cu$_{12}$Sb$_{4}$S$_{13}$ was performed in three steps, according to discussions in Section \ref{sec:fit-in-steps}. First the coefficients $\boldsymbol{\epsilon}$ for the force field potential $E_\text{FF}$ of bonded cation-anion pairs were fitted to high-accuracy DFT force calculations of AIMD snapshots at 800 K.
Fig.~\ref{fig:tetrahedrite-pairpotential} shows the obtained $E_\text{FF}$, including results for two symmetry-inequivalent sub-lattices for both copper and sulfur. 12 Legendre polynomials and a scaling length $r_\text{c}$=1 \AA\ (see Eq.~(\ref{eq:radial-FF})) were used in the fitting.
Secondly, the second and third order residual FCTs $\left(\boldsymbol \psi _2, \boldsymbol \psi _{3}\right)$ were fitted to the residual forces $\mathbf{F} - \mathbb{A}_{\text{FF}} \boldsymbol \epsilon $ of configurations with small random displacement of 0.05 \AA\ on each atom. Thirdly, large displacement (0.3 \AA) and residual forces $\mathbf{F} - \mathbb{A}_{\text{FF}} \boldsymbol \epsilon - \left(\mathbb{A}_2, \mathbb{A}_{3} \right) \left(\boldsymbol \psi _2^T, \boldsymbol \psi _{3}^T\right)^T$ yielded the 4-6th order residual FCTs.
The final LD+FF fitting quality and lattice thermal conductivity results were previously reported \cite{Zhou2014PRL185501} and are not repeated here.

\begin{figure}[htp]
	\includegraphics[width = 0.9 \linewidth]{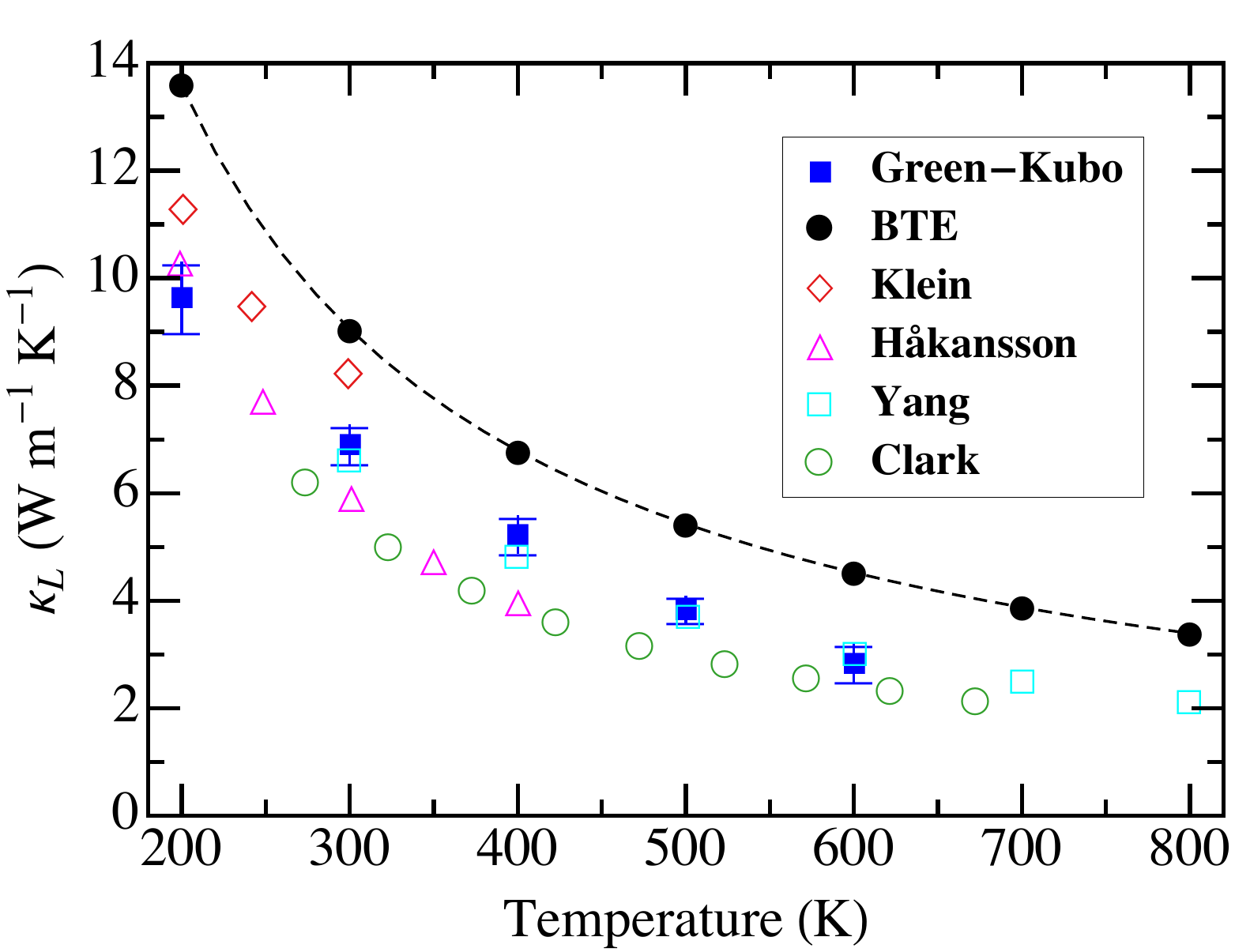}
	\caption{Comparison of the lattice thermal conductivity of NaCl versus temperature from CSLD-based molecular dynamics simulations in conjunction with the Green-Kubo formula (filled squares) third-order PT+BTE calculations (filled circles), and experimental measurements from Refs.~\onlinecite{Hakansson1986JPCS355, Klein1966RSI1291, Clark2013, Yang1981} (open symbols). The dash black line through the PT+BTE results is a guide to the eye. }
	\label{fig:NaCl-kappa}
\end{figure}

NaCl is an interesting case demonstrating the importance of fourth- and higher-order anharmonic interactions for accurate calculations of lattice thermal conductivity. We have conducted a comparative study of its lattice thermal conductivity using PT+BTE and the MD based Green-Kubo linear response formula. Classical MD simulations, based on a fourth-order CSLD fitting, were performed for temperatures between 200 and 600~K, with system sizes ranging from 512 to 4096 atoms. No discernible size-dependence was found in the range of supercells tested. The simulation time varied from 100 ps at 600~K to 1 ns at 200~K, and a timestep of 1 fs was used throughout. At each temperature, a minimum of 10 independent simulations were performed to obtain averages. Fig.~\ref{fig:NaCl-kappa} shows the calculated lattice thermal conductivity values using PT+BTE with 3rd-order terms only and the MD Green-Kubo results with third- and fourth-order anharmonicity. It is seen that the PT+BTE results based on the first-order perturbation theory always overestimate $\kappa_L$ in comparison to the experimental data. We attribute this to the fact that the version of PT+BTE used in our study omits higher-order contributions (e.g., four-phonon processes) to the total scattering rates, and hence overestimates phonon lifetimes and mean free paths. Meanwhile, the Green-Kubo formula gives significantly reduced values of $\kappa_L$, which achieve much better agreement with experiments above the Debye temperature ($\approx$ 300 K for NaCl). This improvement strongly points to the importance of higher-order anharmonicity in NaCl (namely, of fourth-order FCTs). The tendency of the Green-Kubo formula to underestimate $\kappa_L$ at low temperatures (200~K and below) may be attributed to the lack of quantum correction in classical MD, as detailed in Ref.~\onlinecite{Turney2009PRB224305}, although this is also the temperature region where the agreement between the experimental data and PT+BTE is expected to improve. To put our Green-Kubo MD results in a broader perspective, we note that recent theoretical efforts have been devoted to including fourth-order anharmonic contributions to the scattering rates in PT+BTE \cite{Tianli2016,Tianli2017} and to additionally treating temperature-induced anharmonic frequency renormalization \cite{pbte2018,gete2018,Broido2018}. These developments should enable an expanded range of physical accuracy for the computationally efficient PT+BTE framework.

\subsection{Application to type-I Si clathrates}

\begin{figure}[htp]
	\includegraphics[width =1.0 \linewidth]{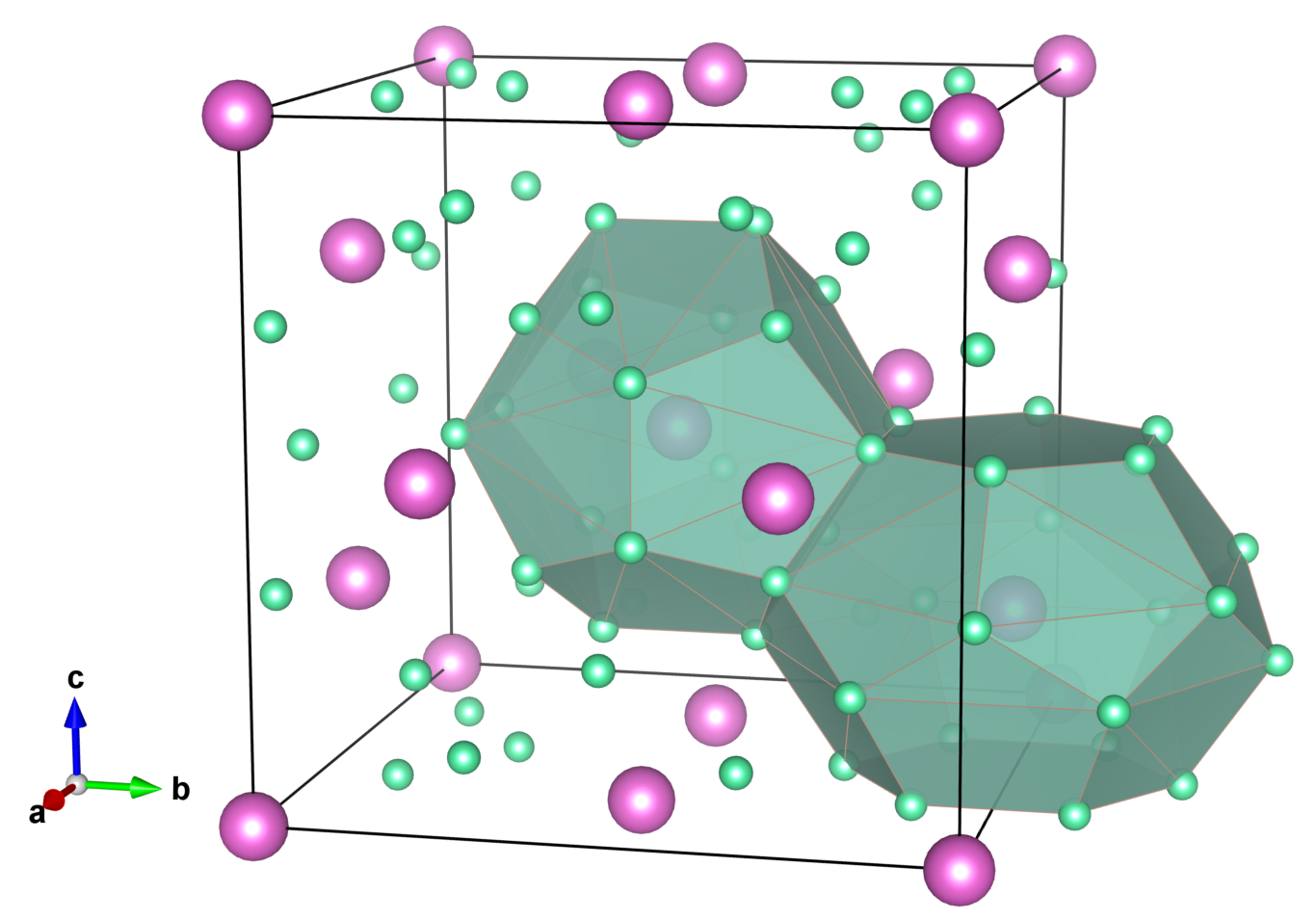}
	\caption{Crystal structure of type-I Si clathrate with guest atoms. Si and guest atoms are displayed as green and magenta solid spheres, respectively. There are two kinds of polyhedra: dodecahedral (D) and tetrakaidecahedral (T).} 
	\label{fig:clathrate-structure}
\end{figure}

Type-I Si clathrate has been investigated as a potential thermoelectric due to its ability to incorporate guest atoms, which significantly lower $\kappa_L$ and donate electrons to the conduction band \cite{Takabatake2014RMP669}. It is often referred to as belonging to the general class of thermoelectric ``phonon-glass electron-crystal'' materials \cite{rowe1995crc}.
Type-I Si clathrate has a complex unit cell with 46 Si atoms in the primitive cell belonging to the space group $Pm\bar{3}n$ \cite{Takabatake2014RMP669} (Fig~\ref{fig:clathrate-structure}). There are two types of empty cages, shaped as dodecahedra (D) and tetrakaidecahedra (T); these cages can be filled with guest atoms (e.g., Na and Ba).
Ongoing research has been focused on clarifying the physical mechanisms by which these guest atoms suppress lattice thermal conductivity \cite{euchner2}. In comparison with other type-I clathrate compounds, such as Ba$_{8}$Ga$_{16}$Sn$_{30}$ and Eu$_{8}$Ga$_{16}$Ge$_{30}$, Si clathrate has smaller cages and contains no disorder in the framework. These factors suppress the tendency towards off-center occupation of guest atoms and make the Si clathrate an ideal platform for examining the specific impact due to guest atoms only.\cite{Takabatake2014RMP669, offcenter} It also represents a good example demonstrating the computational efficiency advantages of CSLD relative to other methods that calculate the anharmonic terms one-by-one. Recently Tadano {it et al} applied CSLD to extract anharmonic force constants of type-I Ge clathrate and studied phonon renormalization due to quartic anharmonicity at finite temperatures \cite{Tadano2018PRL105901}.

\begin{figure}[htp]
		\includegraphics[width =0.8 \linewidth]{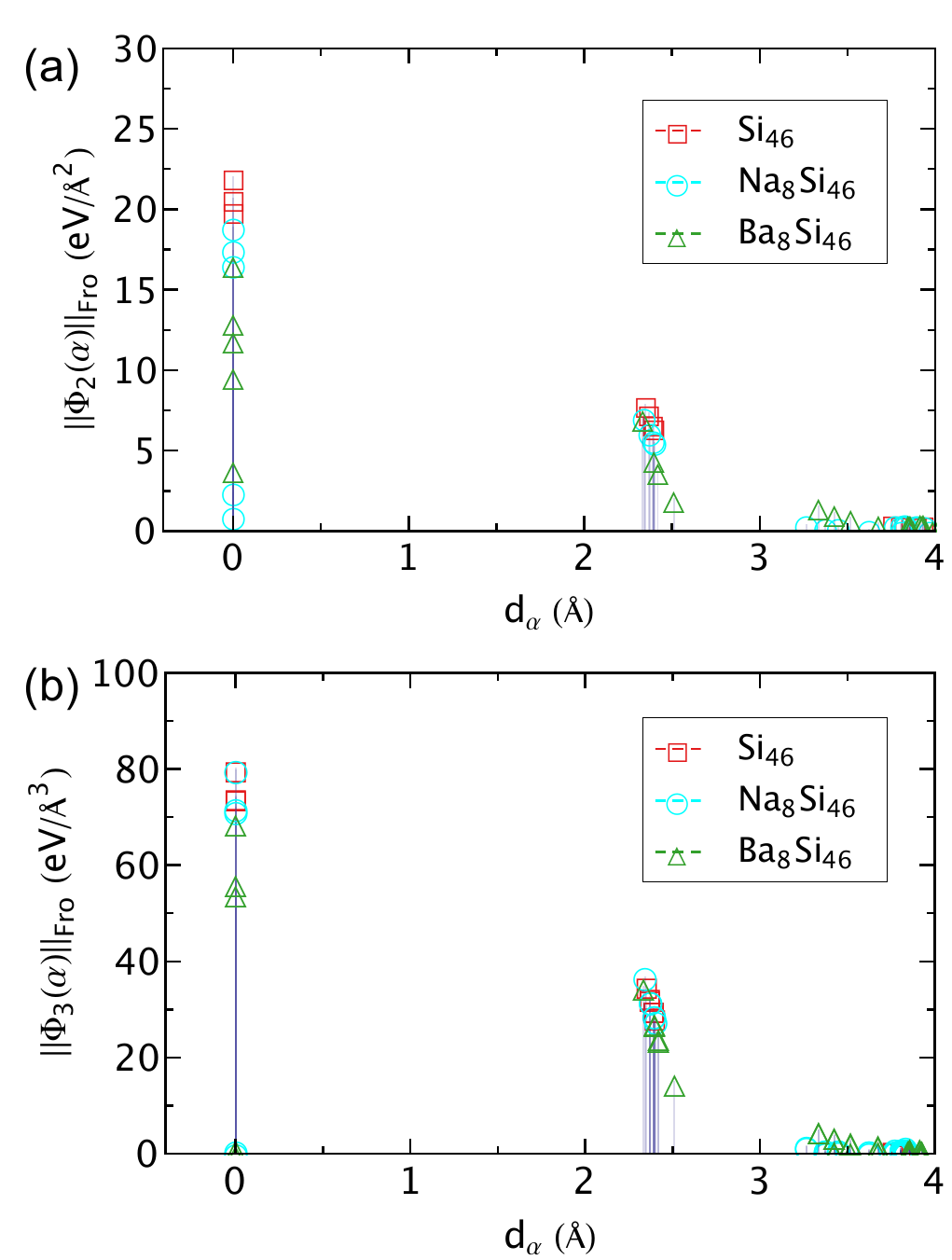}		
		\caption{Frobenius norm of (a) second and (b) third order FCTs versus interaction distance for Si$_{46}$ and $G_{8}$Si$_{46}$ ($G$=Na, Ba).}
		\label{fig:clathrate-FCT-distance}
\end{figure}

Our fully relaxed lattice constants for Si$_{46}$, Na$_{8}$Si$_{46}$ and Ba$_{8}$Si$_{46}$ are 10.229, 10.243, and 10.396 \r{A}, respectively, which compare reasonably well with the experimental values of 10.250, 10.196, and 10.328 \r{A}\cite{si, nasi, basi}. 
 CSLD fitting used twenty 2$\times$2$\times$2 supercell structures for each compound, with all atoms randomly displaced by 0.01-0.04 {\AA}  and including third-order clusters with diameters not exceeding the second-nearest neighbor shell. We note that almost 600 calculations would be necessary to calculate all the included FCTs in a one-by-one manner, such as currently implemented in Refs.~\onlinecite{shengbte,shengbte3rd,shengbteGau,phono3py}.

\begin{figure}[htp]
	\includegraphics[width =0.8 \linewidth]{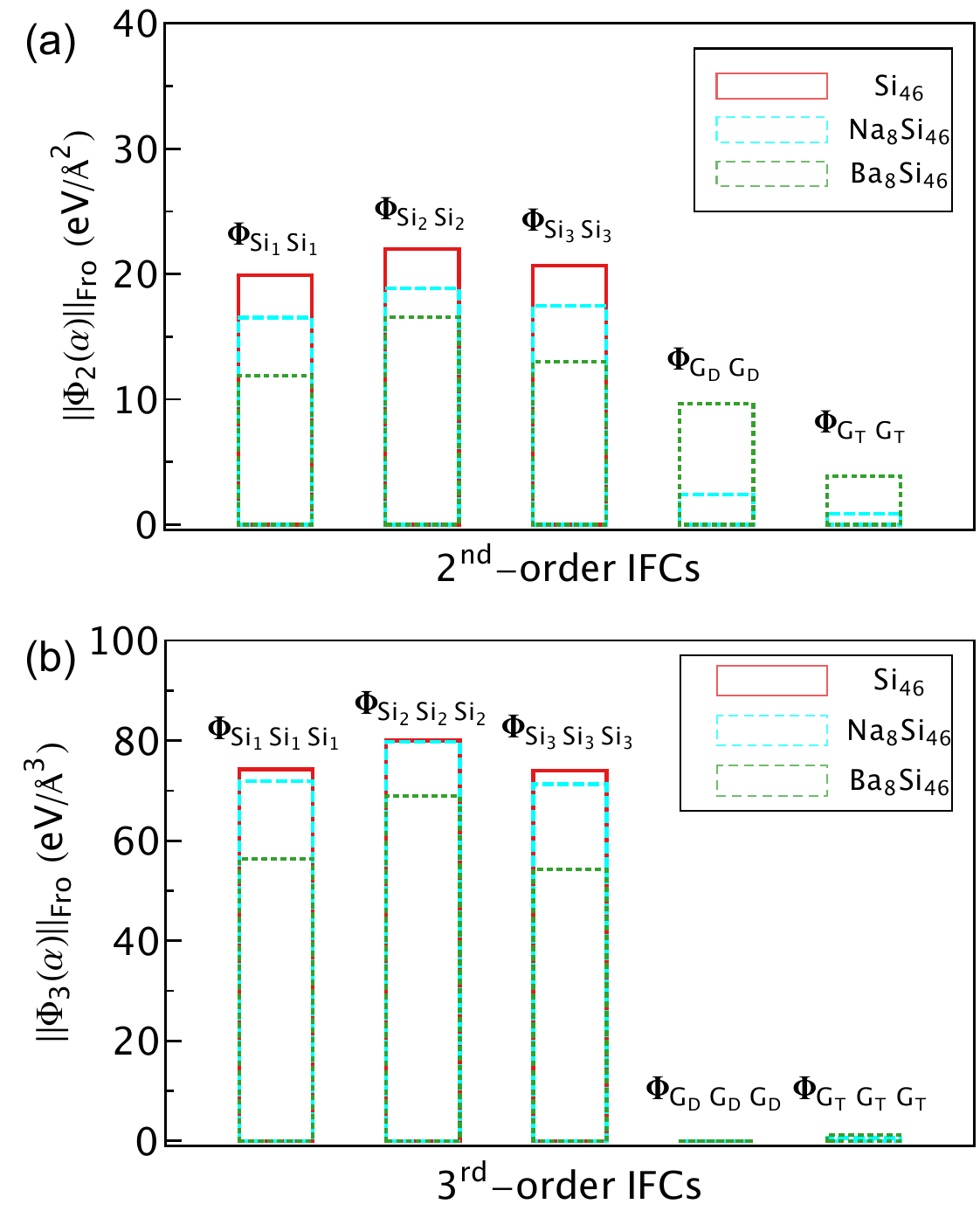}		
	\caption{Comparison of Frobenius norm of (a) the improper on-site pair FCT $\mathbf{\Phi}_{aa}$ and (b) improper on-site third order $\mathbf{\Phi}{aaa}$ for Si$_{46}$ and $G_{8}$Si$_{46}$ ($G$=Na, Ba). The subscripts on $G$ refer to the two types of cages (dodecahedral and tetrakaidecahedral).}
	\label{fig:clathrate-FCT}
\end{figure}

As expected, the calculated FCTs for both the second and third order clusters are found to be dominated by short-range  (up to the first-coordination shell) interactions, as can be seen from Fig.~\ref{fig:clathrate-FCT-distance}.
We regard the improper on-site pair $\mathbf{\Phi}_{aa}$ as a measure of the overall strength of the harmonic forces experienced by $a$, since $\mathbf{\Phi}_{aa}$ is the negative of the sum of all pair interactions according to the ASR in Eq.~(\ref{eq:ACsumrule}).
Figure~\ref{fig:clathrate-FCT}(a) compares $\mathbf{\Phi}_{aa}$ for Si$_{46}$, Na$_{8}$Si$_{46}$, and Ba$_{8}$Si$_{46}$. We observe that (1) the addition of guest atoms weakens $\mathbf{\Phi}_{\text{SiSi}}$, the effect being more pronounced for the larger Ba atoms, (2) the one-body restoring forces on the guest atoms, $\mathbf{\Phi}_{GG}$, are  much weaker than those on the host Si atoms, (3) $\mathbf{\Phi}_{GG}$ of the guest atoms in the dodecahedral (D) cages are stronger than those in tetrakaidecahedral (T) cages, and (4)  the interaction of the Ba atoms with the surrounding Si cage is considerably stronger than that of the Na atoms.

The third-order on-site FCT $\mathbf{\Phi}_{aaa}$ may serve as a measure of the strength of anharmonic interactions according to the generalized ASR in Eq.~(\ref{eq:translational-invariance-orbit}).  Figure~\ref{fig:clathrate-FCT}(b) shows that $\mathbf{\Phi}_{aaa}$ of the guest atoms are zero by symmetry in the D cages and very small in the T cages. We also point out that Na guest atoms have little impact on the strength of the anharmonic Si terms, $\mathbf{\Phi}_{\text{SiSiSi}}$, while Ba guest atoms lead to an appreciable decrease of up to a few tens of percent. 
This analysis shows that Na behaves more like a perfect ``rattler'' which has relatively weak influence on the bond strength of the Si framework, while Ba has stronger interactions with the host and considerably reduces both the stiffness and anharmonicity of the Si-Si bonds.

\begin{figure*}[htp]
	\includegraphics[width = 1.0 \linewidth]{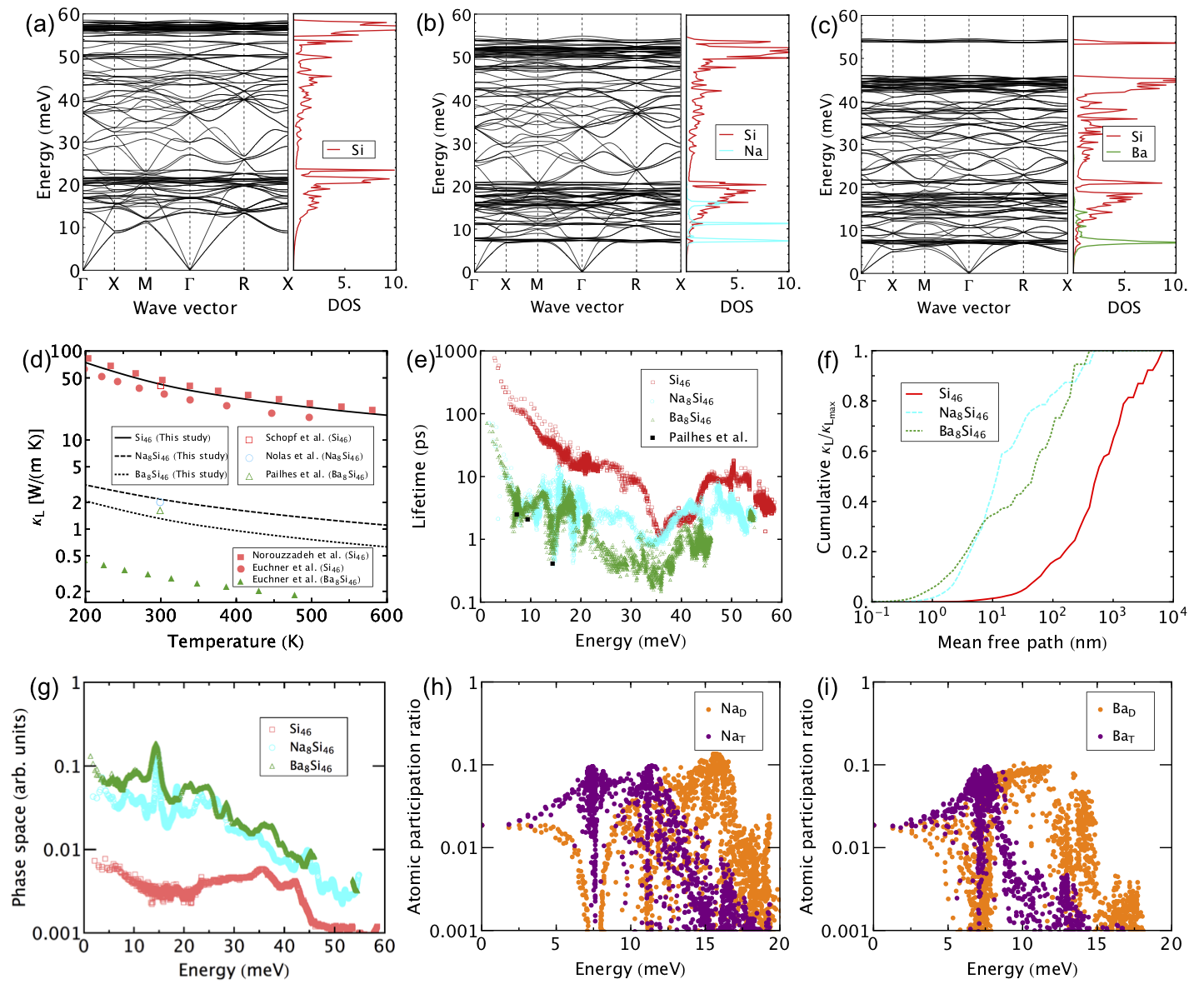}
	\caption{Calculated phonon dispersions for (a) Si$_{46}$, (b) Na$_{8}$Si$_{46}$ and (c) Ba$_{8}$Si$_{46}$.  (d) Calculated lattice thermal conductivities compared to previous theoretical study using effective potential by Schopf{\it\ et al.\/} \cite{schopf} (red square), first-principles calculations by Norouzzadeh{\it\ et al.\/}~\cite{Norouzzadeh2017} (red squares) and Euchner{\it\ et al.\/}~\cite{Euchner2018} (red disks and green triangles),  and experimental measurements by Pailh\`es{\it\ et al.\/}\cite{euchner2} (green triangle) and Nolas {\it et al.\/}\cite{nolas} (blue circle). (e) Calculated mode-resolved lifetimes at 300~K compared with experimental measurements performed on Ba$_{8}$Si$_{46}$ \cite{euchner2}. (f) Normalized cumulative lattice thermal conductivities at 300~K.  (g) Comparison of three-phonon scattering phase space in Si$_{46}$ and $G_{8}$Si$_{46}$ ($G$=Na, Ba). The three-phonon scattering phase space is calculated using $\Gamma_{\lambda\lambda^{\prime}\lambda^{\prime\prime}}^{\pm}$ by setting $ |V_{\pm\lambda\lambda^{\prime}\lambda^{\prime\prime}}^{(3)}| $ as a constant.  Atomic partition ratio of (h) Na and (i) Ba atoms. 
	}
	\label{fig:clathrate}
\end{figure*}

The calculated phonon dispersion curves and atom-projected phonon densities-of-states (DOS) are shown in Fig.~\ref{fig:clathrate}(a-c). The incorporation of guest atoms in the empty cages slightly suppresses the Debye temperatures due to the softening of Si-Si interactions. Pronounced change of the vibrational spectrum is found in the low-frequency window from 0 to 20~meV. In contrast to Si$_{46}$, three Einstein-like frequency peaks associated with Na atoms are centered at energies $E_{1}\approx$ 6-8 meV, $E_{2}\approx$ 10-12 meV and $E_{3}\approx$ 14-16 meV. The dispersion of these optical modes associated with the Na atoms is rather flat, cutting through the acoustic branches and thus leading to an ``avoided crossing'' behavior, which directly decreases the group velocities of acoustic phonons. The resulting strongly localized modes can be attributed to the weak interaction between guest (Na) and host (Si) atoms, as seen in Fig.~\ref{fig:clathrate-FCT}(a) (see also Ref.~\onlinecite{toberer}). 
We find that Ba atoms in Ba$_{8}$Si$_{46}$ display similar characteristics in phonon DOS as identified for Na atoms. The slight difference lies in the fact that the Ba contribution to the phonon DOS seems to have a relatively extended distribution with only one broadened frequency peak resembling a single-frequency Einstein model at energy $E_{1}^{\prime}\approx$ 6-8 meV, corresponding to the vibrations of Ba atoms in T cages. This raises the question whether or not there is essential difference in localization levels between Na and Ba atoms, which motivates us to explicitly calculate the atomic partition ratio using the definition in Ref.~\onlinecite{tadano}. As shown in Fig.~\ref{fig:clathrate}(h) and (i), Na and Ba display very similar localization levels, while the difference in their contributions to phonon DOS probably can be attributed to the differences in their atomic mass and size.

The computed lattice thermal conductivities are shown in Fig.~\ref{fig:clathrate}(d). Our results agree reasonably well with experimental measurements \cite{euchner2,nolas} and with the values obtained by other computational studies using either effective potential~\cite{schopf} or first-principles calculations.\cite{Norouzzadeh2017,Euchner2018} We note that although our calculated $\kappa_{L}$ of \ce{Ba8Si46} is considerably larger than those reported by Euchner{\it\ et al.\/}~\cite{Euchner2018}, both of the results confirm the significantly reduced $\kappa_{L}$ in \ce{Ba8Si46} compared to Si$_{46}$. The calculated phonon lifetimes in Fig.~\ref{fig:clathrate}(e) are in good agreement with experimental measurements for \ce{Ba8Si46}\cite{euchner2}, and they 
further elucidate the physics of the  lattice thermal conductivity suppression due to the presence of guest atoms. 
In particular, we find that the phonon lifetimes are greatly shortened in \ce{Na8Si46} and \ce{Ba8Si46}, especially near the rattling modes.   This seemingly contradicts to the observed weakened third-order anharmonicity. However, phonon lifetimes may also be significantly altered due to the change in phonon dispersions, which motivates us to examine the impact of three-phonon scattering phase space that excludes the effect of third-order anharmoncity. We find that, as shown in Fig.~\ref{fig:clathrate}(g), the addition of Na and Ba guest atoms drastically increases the three-phonon scattering phase space by nearly an order of magnitude, thus explaining the significantly reduced lifetimes. 
As a result of the reduction in both phonon lifetime and group velocity, the phonon mean free paths (MFPs) are much smaller in the filled clathrates \ce{Na8Si46} and \ce{Ba8Si46}, which is illustrated in Fig.~\ref{fig:clathrate}(f). Interestingly, our computations show that there is a considerable contribution to $\kappa_L$ from acoustic phonon modes with MFPs longer than 20\r{A}, thus questioning the common assumption that the MFP is limited by the separation of the guest atoms acting as scattering centers.\cite{christensen} This observation also suggests that it may be possible to further reduce $\kappa_L$ by introducing atomic disorder in the framework to scatter these long-MFP phonons.

\section{Conclusion} \label{sec:conclusion}
We have described an anharmonic lattice dynamics model in a cluster-based form that is convenient for keeping track of high order force constants and a compressive sensing framework tailored towards force constant extraction from first-principles calculations. CSLD is general, conceptually straight-forward and computationally much more efficient than other currently existing methods for treating anharmonicity. For instance, CSLD can easily include 4th- to 6th-order anharmonic terms, which are currently inaccessible to DFPT and ``2n+1'' methods. Such terms are particularly important for compounds with double-well type potentials for harmonically unstable modes, such as thermoelectric \ce{Cu12Sb4S13} and ferroelectric perovskites (e.g., \ce{BaTiO3}), and we expect that CSLD will contribute to advancing our theoretical understanding of these fascinating systems. 
The accuracy of the lattice dynamics model can be improved systematically by simply increasing the size of the training set, even though some inherent difficulties associated with the Taylor expansion basis set exist, including the growing numeric instability and error propagation associated with high-order polynomials, and the difficulty of treating the highly anharmonic potential within the first-coordination shell. The introduction of pairwise potentials for covalently bonded atoms into the CSLD formalism has shown to be effective in combating some of these difficulties. The software package {\sc CSLD} is now freely available at \url{https://github.com/LLNL/csld}. Applications of CSLD for phonon calculations will be presented in Part II of the series.

\begin{acknowledgments}
The work of F.Z.\ was supported by the Critical Materials Institute, an Energy Innovation Hub funded by the U.S. Department of Energy, Office of Energy Efficiency and Renewable Energy, Advanced Manufacturing Office, and performed under the auspices of the U.S. Department of Energy by LLNL under Contract DE-AC52-07NA27344. Y.X., W.N., and V.O. acknowledge support from the Office of Naval Research under award No. N00014-14-1-0444, U.S. Department of Energy, Office of Science, Basic Energy Sciences, under Grant DE-FG02-07ER46433, and computational resources from the National Energy Research Scientific Computing Center, which is supported by the Office of Science of the U.S. Department of Energy under Contract No. DE-AC02-05CH11231.\end{acknowledgments}

\appendix
\renewcommand\thefigure{\Alph\arabic{figure}}    
\setcounter{figure}{0}    
\numberwithin{equation}{section}
\renewcommand{\thesection}{\Alph{section}}

\section{Multi-index notation} \label{sec:multi-index}
In the multi-index notation for $N$ variables, an $N$-tuple $\alpha=(\alpha_{1}, \dots, \alpha_{N})$ of non-negative integers is defined with the following notations for its absolute value, factorial, power, and partial derivative, respectively:
\begin{eqnarray}
|\alpha| &=& \sum_{a} \alpha_{a} \\
\alpha ! &=& \prod_{a} \alpha_{a}! \\
u^{\alpha} &=& \prod_{a} u_{a}^{\alpha_{a}} \label{eq:multi-index-ualpha} \\
\partial_{ u^{\alpha}} &=&  \partial^{|\alpha|}/ \prod_{a}\partial u_{a}^{\alpha_{a}}.
\label{eq:multi-index}  
\end{eqnarray}
 The $N$-tuple $\alpha$ is equivalent to a ``flattened'' list (hereby called{ \bf cluster}) $C(\alpha)$ of length $|\alpha|$
$$
C(\alpha)= \{ a,  \dots, a, a, \dots \}  
$$
where element $a$ appears exactly $\alpha_{a}$ times. Hereafter we refer to the $N$-tuple $\alpha$ and the corresponding $|\alpha|$-element cluster $C(\alpha)$ indistinguishably.  
In a {\bf proper} cluster $\alpha_a \leq 1$.

In this work, the $N$-tuple is used for designating a cluster of atoms or lattice sites.   When used in conjunction with another list of index $I = \{i\}$, we define
$$
u^{\alpha}_I = \prod_{k=1}^{|\alpha|} u_{a_k,i_k}
$$
For example, given a proper pair cluster composed of sites $s_1$ and $s_2$: $\alpha = (1, 1, 0, \dots) = \{s_1, s_2\}$, $\alpha! = 1$, $\Phi_{I} (\alpha) = \Phi_{1 i_{1}, 2 i_{2}}$, and $u^{\alpha}_{I} = u_{1 i_{1}} u_{2 i_{2}}$.
Given an improper triplet cluster of site $s_1$ alone: $\alpha = (3, 0, \dots)=\{s_1, s_1, s_1\}$, $\alpha! = 6$, $\Phi_{I} (\alpha) = \Phi_{1 i_{1}, 1 i_{2}, 1 i_{3}}$, and $u^{\alpha}_{I} = u_{1 i_{1}} u_{1 i_{2}} u_{1 i_{3}}$.
According to above definitions the potential expansion of Eq.~(\ref{eq:Taylor}) can be conveniently translated into Eq.~(\ref{eq:expansion-multi-index}).

\section{Iterative null-space construction} \label{sec:null-space}
Given a set of $N_c$ linear constraints on the $N_v$-dimensional variable $\mathbf{\Phi}$ such that certain linear combinations of $\mathbf{\Phi}$ vanish:
$$
\mathbb{B}_{k} \mathbf{\Phi} =0,
$$
where each $\mathbb{B}_{k}$ is an $n_k \times N_v$ matrix, $k = 1 \dots N_c$, the number of independent variables is reduced from $N_v$ to the dimension of the (right) null space of the full constraint matrix $\mathbb{B}$ composed of all constraint matrices $\mathbb{B}^T = (\mathbb{B}_{1}^T, \dots, \mathbb{B}_{N_c}^T)$, and the independent variables may be chosen according to the basis vectors $\{\mathbf{c}_1, \dots \}$ of the null space. An iterative procedure to solve for these basis vectors in columns $\mathbb{C} =\left(\mathbf{c}_1, \dots\right)$ is:
\begin{enumerate}
\item Initialize $\mathbb{C} = \mathbbm{1}_{N_v}$, tolerance $\delta$ (e.g.\  $10^{-8}$) 
\item For $k$ in $1 $--$ N_c$
\begin{enumerate}
\item Row-reduce  $\mathbb{B}_k \mathbb{C}$ by Gauss-Jordan elimination into a row echelon form to find its null space basis vectors in columns, $\mathbb{C}'$. To enhance numerical stability, elements of $\mathbb{C}'$ with absolute values below $\delta$ are set to 0.
\item Update $\mathbb{C} \leftarrow \mathbb{C} \mathbb{C'} $.
\end{enumerate}
\item Return $\mathbb{C} $.
\end{enumerate}

\end{document}